\documentclass[sigconf]{acmart}

\usepackage{booktabs} 
\usepackage[flushleft]{threeparttable}
\usepackage{xcolor}
\usepackage[utf8]{inputenc}
\usepackage{float}
\usepackage{amssymb}
\usepackage{ifthen}
\usepackage{multirow}
\usepackage{caption}
\usepackage{listings}
\usepackage{enumitem}
\usepackage{array,graphicx}
\usepackage{soul}
\usepackage{balance}
\usepackage{pifont}
\usepackage{etoolbox,siunitx}
\usepackage{nicefrac}
\usepackage{mathtools}

\usepackage[labelsep=quad,indention=10pt]{subfig}
\usepackage{lipsum} 

\usepackage{tikz}
\usepackage{pgfplots}
\usepgfplotslibrary{statistics}
\usetikzlibrary{fadings}
\usetikzlibrary{arrows.meta,shapes.geometric,shadows}

\usepackage{xcolor,pifont}
\newcommand*\colourcheck[1]{%
  \expandafter\newcommand\csname #1check\endcsname{\textcolor{#1}{\ding{52}}}%
}
\newcommand*\colourcross[1]{%
  \expandafter\newcommand\csname #1cross\endcsname{\textcolor{#1}{\ding{56}}}%
}

\usepackage[lined,boxruled,norelsize,linesnumbered]{algorithm2e}
\def\HiLi{\leavevmode\rlap{\hbox to \hsize{\color{gray!35}\leaders\hrule height .8\baselineskip depth .5ex\hfill}}}

\newtheorem{defn}{Definition}

\definecolor{lightgray}{gray}{0.85}
\definecolor{bananayellow}{rgb}{1.0, 0.88, 0.21}

\usepackage{graphicx}
\usepackage{seqsplit}

\newcommand{\tool}{\textsc{SelfOracle}\xspace} %
\newcommand{\deeproad}{DeepRoad\xspace} %

\SetKwComment{Comment}{$\triangleright$\ }{}
\SetCommentSty{itshape}
\newboolean{showcomments}
\setboolean{showcomments}{true}
\ifthenelse{\boolean{showcomments}}
{\newcommand{\nb}[2] {
  \fcolorbox{black}{gray!20}{\bfseries\sffamily\scriptsize#1:}
  {\sf\small$\blacktriangleright$\textit{#2}$\blacktriangleleft$}
}
}
{\newcommand{\nb}[2]{}
}

\newcommand{\head}[1]{\noindent\textbf{#1.}}

\newcounter{fcounter}
\newcommand{\curl}[1]{\footnote{\url{#1}}}

\makeatletter
\newcommand{\thickhline}{%
    \noalign {\ifnum 0=`}\fi \hrule height 1pt
    \futurelet \reserved@a \@xhline
}
\makeatother

\renewcommand\footnotetextcopyrightpermission[1]{} 

\setcopyright{none}

\begin{document}
	 

\title{Misbehaviour Prediction for Autonomous Driving Systems}


\author{Andrea Stocco}
\orcid{0000-0001-8956-3894}
\affiliation{
  \institution{Universit\`a della Svizzera Italiana}
  \city{Lugano}
  \country{Switzerland}
}
\email{andrea.stocco@usi.ch}

\author{Michael Weiss}
\affiliation{
  \institution{Universit\`a della Svizzera Italiana}
  \city{Lugano}
  \country{Switzerland}
}
\email{michael.weiss@usi.ch}

\author{Marco Calzana}
\affiliation{
  \institution{Universit\`a della Svizzera Italiana}
  \city{Lugano}
  \country{Switzerland}
}
\email{marco.calzana@usi.ch}

\author{Paolo Tonella}
\affiliation{
  \institution{Universit\`a della Svizzera Italiana}
  \city{Lugano}
  \country{Switzerland} 
}
\email{paolo.tonella@usi.ch}


\begin{abstract}
Deep Neural Networks (DNNs) are the core component of modern autonomous driving systems. 
To date, it is still unrealistic that a DNN will generalize correctly in all driving conditions. Current testing techniques consist of offline solutions that identify adversarial or corner cases for improving the training phase, and little has been done for enabling online healing of DNN-based vehicles. 

In this paper, we address the problem of estimating the confidence of  DNNs in response to unexpected execution contexts with the purpose of predicting potential safety-critical misbehaviours such as out of bound episodes or collisions. 
Our approach \tool is based on a novel concept of self-assessment oracle, which monitors  the DNN confidence at runtime, to predict unsupported driving scenarios in advance.
\tool uses autoencoder and time-series-based anomaly detection to reconstruct the driving scenarios seen by the car, and 
determine the confidence boundary of normal/unsupported conditions. 

In our empirical assessment, we evaluated the effectiveness of different  variants of \tool at predicting injected anomalous driving contexts, using DNN models and simulation environment from Udacity. Results show that, overall, \tool can predict 77\% misbehaviours, up to 6 seconds in advance, outperforming the online input validation approach of DeepRoad by a factor almost equal to 3.

\end{abstract}

%
%
%



\maketitle

\section{Introduction}\label{sec:introduction}

\pagestyle{plain} 

Self-driving cars are one of the emerging technologies nowadays, and possibly the standard way of transportation in the future. 
Such autonomous driving systems receive data from a multitude of sensors, and analyze them in real time using deep neural networks (DNNs) to determine the driving parameters for the actuators.

To test such complex software systems, companies perform a limited number of expensive in-field tests, driving a car on real world streets, or within closed-course testing facilities~\cite{Cerf:2018:CSC:3181977.3177753}. This provides detailed sensor data of the vehicle that are recorded, played back, and recreated within a simulator to obtain comprehensive test scenarios.
Simulation-based test scenarios allow re-testing new autopilot releases on a large numbers of nominal conditions, as well as challenging (e.g., adverse weather) and dangerous circumstances (e.g., a pedestrian suddenly crossing the road), at a low cost~\cite{Cerf:2018:CSC:3181977.3177753}.


The potentially unlimited number of testable driving scenarios, combined with the lack of human interpretability of the internal functioning of DNNs~\cite{AlvarezMelisJ18}, makes it difficult to predict the vehicle's (mis-)behaviour with respect to unforeseen edge-case scenarios. 
Misbehaviours span a wide range of situations, associated with different degrees of severity, from cases where the car does not drive smoothly (e.g., excessively high derivative of the steering angle over time), to safety-critical failures and casualties~\cite{tesla-accident-1,tesla-accident-2,tesla-accident-3,tesla-accident-4}.
In this respect, promptly detecting unexpected, untested behaviours is of paramount importance, so as to make sure that human-driven or self-healing corrective actions take place to ensure safety.

In this paper, we tackle the \textit{self-assessment oracle problem} for autonomous driving system, i.e., the problem of monitoring the confidence level of a DNN-based autonomous driving system in order to timely predict the occurrence of future misbehaviours. 
The problem is critical because a failure in detecting an unexpected condition may have severe consequences (i.e., a fatal crash), whereas false alarms, even if not dangerous, may cause driver's discomfort and negatively affect the driving experience. Creating a self-assessment oracle that evaluates the confidence of a DNN \textit{at runtime}, and predicts whether the system is within a low-confidence zone, is a largely unexplored research problem.

Challenges arise because unexpected driving conditions are, by definition, unknown at training time, otherwise they would be used to train a better DNN~\cite{Campos:2016:EUO:2962863.2962870}. As a consequence, the problem being addressed belongs to the unsupervised class of data analysis,  
and we have to infer the unexpected only by looking at the normal driving scenarios.
Moreover, the ensemble of possible misbehaviours for a DNN-based system is vast and necessarily domain-dependent, being associated with deviations from the functional requirements. 
In recent years, researchers have proposed solutions for testing autonomous driving systems software~\cite{deepsafe,Gambi:2019:ATS:3293882.3330566,deepxplore,deeptest,deeproad,Abdessalem-ASE18-1,Abdessalem-ASE18-2,Abdessalem-ICSE18,deepsafe}. 
A number of approaches propose input generation techniques that produce corner/adversarial cases, used to improve the robustness of self-driving car modules by re-training~\cite{deepxplore,deeptest,deeproad,Kim:2019:GDL:3339505.3339634}. Other works target test case generation to expose faults for extreme conditions, such as the vehicle colliding with a pedestrian~\cite{Abdessalem-ICSE18}, or driving off the road~\cite{Gambi:2019:ATS:3293882.3330566}.
All these approaches concern \textit{offline} solutions for improving the robustness and reliability of DNNs, achieved by enhancing the training data and the autopilot module, which are extended to include underrepresented critical scenarios. 
To the best of our knowledge, no existing work is focused on online confidence estimation for self-driving vehicles with  
the aim of anticipating the occurrence of misbehaviours so as to enable self-healing procedures that can avoid future failures. 

In this paper, we propose a novel self-assessment oracle for autonomous vehicles based on \textit{confidence estimation}, \textit{probability distribution fitting}, and \textit{time series analysis}. 
Our technique is implemented in a tool called \tool, which leverages reconstruction-based techniques from the deep learning (DL) field to analyze spatiotemporal historical driving information. The reconstruction error is used as a black-box confidence estimation for the DNN. 
During probability distribution fitting, \tool captures the behaviour of the self-driving car under nominal conditions, and fits a Gamma distribution to the observed data. Analytical knowledge of Gamma's parameters allows \tool to estimate an optimal confidence threshold, as a trade-off between prediction of all misbehaviours (the stricter the threshold, the better), and probability of false alarms (reduced when threshold is higher).
By observing a decreasing confidence trend over time, \tool can anticipate a misbehaviour 
by recognizing unexpected conditions timely enough to enable healing actions (e.g., manual or automated disengagement). 

We have evaluated the effectiveness of \tool on the Udacity simulator of self-driving cars~\cite{udacity-simulator}, using DNNs available from the literature. 
We have modified the simulator to being able to inject unexpected driving conditions in a controllable way (i.e., day/night cycle, rain, or snow). Such injected conditions are by construction disjoint from those used to train the DNNs. 
In our experiments on 72 simulations, 
\tool is able to safely anticipate 77\% out of bound episodes/crashes, up to 6 seconds in advance. The corresponding false alarm rate in nominal driving conditions is 1\%. A comparative experiment with the online input validation strategy of DeepRoad~\cite{deeproad} shows that \tool achieves substantially superior misbehaviour prediction on all the considered effectiveness metrics (e.g., twice as high in terms of AUC-PR and almost three times higher in terms of FPR). 

Our paper makes the following contributions:

\begin{description}[noitemsep]
\item [Technique] An unsupervised technique for misbehaviour prediction based on confidence estimation, probability distribution fitting and time series analysis, implemented in the tool \tool, which is available~\cite{tool}.
\item [Simulator] An extension of the Udacity simulator to inject unexpected driving conditions dynamically during the simulation. 
\item [Evaluation] An empirical study  showing that the reconstruction error used by \tool for time series analysis is a promising confidence metric for misbehaviour prediction, outperforming the online input validation approach of \deeproad.
\item [Dataset] A dataset of 765 labeled simulation-based collision and out-of-bound episodes that can be used to evaluate the performance of prediction systems for autonomous driving cars.
\end{description}


\section{Background}\label{sec:background}

\head{DNN-based Autonomous Vehicles}
Self-driving cars (SDC, hereafter) have benefited from many technological advancements both in hardware and in software. 
Data gathered by LIDAR sensors, cameras, and GPS are analyzed in real time by advanced Deep Neural Networks (DNN) which govern over the actual maneuvers of the car (i.e., steering, braking, acceleration).
In order to manage a wide variety of driving scenarios, SDCs necessitate a large amount of driving data, combining nominal and adversarial scenarios~\cite{10-million-miles}. 

To date, it is still unlikely for a DNN to generalize correctly to the plethora of driving situations met everyday by human drivers.
As such, a component monitoring the \textit{confidence} of the DNN may promptly detect when the SDC is entering a low-confidence zone, and activate healing strategies that bring the vehicle to a safe state.

In self-driving cars, depending on the level of autonomy, the self-healing procedure can either involve the human driver, or can be delegated to an automated system.\footnote{https://www.nhtsa.gov/technology-innovation/automated-vehicles-safety} 
At both levels, early and accurate misbehaviour prediction is an essential precondition to enable safe healing, and an overall pleasant driving experience.

\head{Confidence Measures in DNNs}
The prediction must consider DNN uncertainties originating from the measurements in response to possible adverse environmental conditions in which the SDC operates. 
The confidence level of a SDC can be measured through white-box or black-box techniques. 
White-box monitoring of the internal behaviour of a DNN component. For simple classifiers, measuring softmax probabilities, or information theoretic metrics such as entropy, and mutual information~\cite{Shannon1948} may suffice. For more complex networks such as those than operate on a SDC, softmax probabilities and entropy are known to be unreliable confidence estimators~\cite{8683359}. Moreover, white-box metrics require a transparent access to the network, and substantial domain-knowledge for the creation of nontrivial probabilistic models that approximate the network's uncertainty.

Black-box techniques, differently, model the SDC uncertainty by monitoring the relation between the current input (images) and the input data used during training. For instance, consider a SDC which has been trained only with images representing highways. If images representing a narrow city street are given to the DNN, the model will still output steering angles, but ideally we would like to warn the SDC of a drop in the confidence level. The main advantages of black-box confidence metrics consist in being independent from the specific SDC architecture, requiring no modifications to the existing DNN model because they use information which is readily available for analysis, and in being, therefore, highly generalizable. 
In this paper, we focus on black-box confidence estimation, because softmax probabilities and entropy are known to be unreliable confidence estimators to model complex DNN networks~\cite{8683359}. 
We next describe autoencoders and time series analysis, which are the main building blocks of our approach. 

\begin{figure*}[t]
\centering
\includegraphics[trim=0cm 10cm 1cm 0cm, clip=true,width=0.8\textwidth]{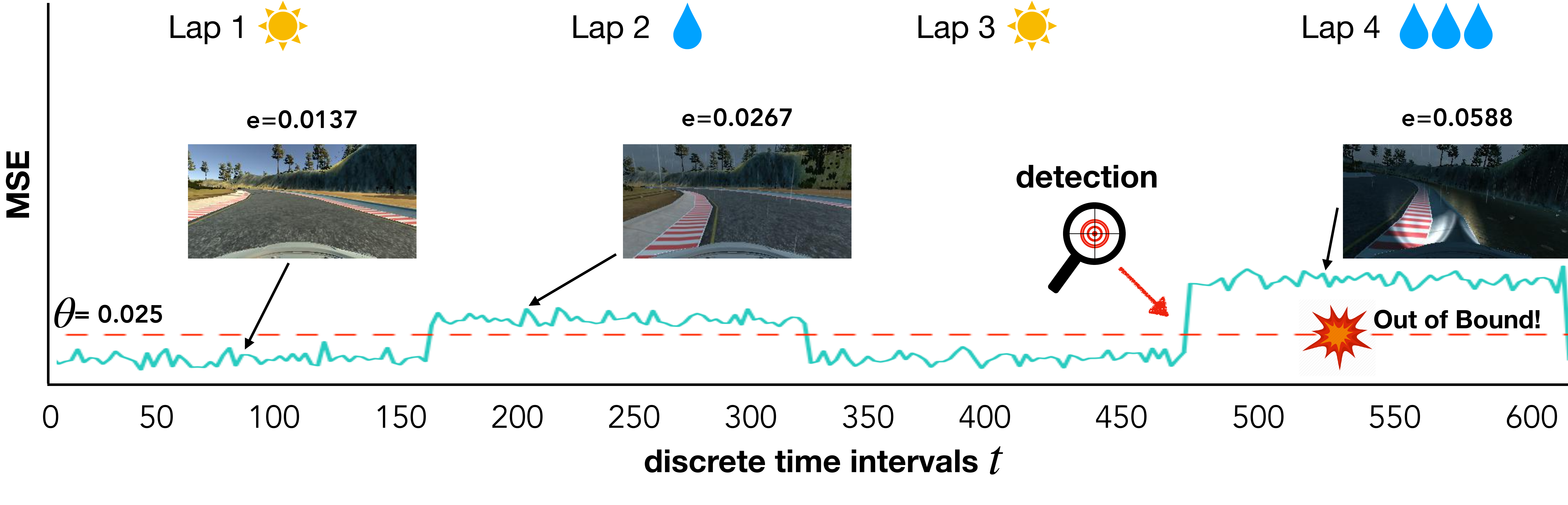}
\caption{Confidence levels of NVIDIA's DAVE-2~\cite{nvidia-dave2} in response to changing driving scenarios. 
The picture shows frames captured during the execution of the SDC along one of our testing tracks, under different conditions: (Lap~1)~sun, (Lap~2)~light rain, (Lap~3)~sun, and (Lap~4)~heavy rain.
The picture juxtaposes the reconstruction error by the anomaly detector of \tool, which is used as a proxy for the DNN confidence.
We can notice that the reconstruction error is low when the car drives under sunny conditions (i.e., conditions similar to those observed during training), whereas the error increases moderately with adverse conditions (the car does no longer follow the center of the road), and grows above a given threshold when facing heavy rainy conditions at night time (which cause the SDC to drive off the road).} 
\label{fig:example} 
\end{figure*}

\head{Autoencoders}\label{sec:autoencoders}
An autoencoder (AE) is a DNN designed to reconstruct its own input. It consists of two sequentially connected components (an encoder, and a decoder) that are arranged symmetrically. 
The simplest form of autoencoder (SAE) is a three-layer DNN: the input layer, the hidden layer, and the output layer.
The hidden layer encodes any given input $\boldsymbol{x} \in \mathcal{R}^{D}$  to its internal representation (\textit{code}) $\boldsymbol{z} \in \mathcal{R}^{Z}$ with a function
$f(\boldsymbol{x}) = \boldsymbol{z}$. Usually $Z \ll D$.
The output layer (decoder) decodes the encoded input with a reconstruction function $g(\boldsymbol{z}) = \boldsymbol{x'}$, where $\boldsymbol{x'}$ is the reconstructed input $\boldsymbol{x}$. 
The autoencoder minimises a loss function $\mathcal{L}(\boldsymbol{x}, g(f(\boldsymbol{x})))$, which measures the distance between the original data and its low-dimensional reconstruction. A widely used loss function in autoencoders is the Mean Squared Error (MSE).



The input and output layers of autoencoders have the same number of nodes. If multiple hidden layers are used, the architecture is referred to as deep autoencoder (DAE). 
The surge of novel kinds of DNNs has correspondingly produced variants of autoencoders based on such architectures. For example, 
convolutional autoencoders~\cite{10.1007/978-3-642-21735-7_7}
allow learning powerful spatial-preserving relationship within images at a lower training time with respect to fully dense layers.
Another interesting proposal are variational autoencoders (VAE) that are able to model the relationship between the latent variable $\boldsymbol{z}$ and the input variable $\boldsymbol{x}$ by learning the underlying probability distribution of observations using variational inference~\cite{An2015VariationalAB}.

\head{Time Series Analysis}\label{sec:rnn}
Traditional feedforward DNNs assume that all inputs and outputs are independent of each other. However, learning temporal dependencies between inputs or outputs is important in tasks involving continuous streams of data. 
Thus, a predictive model can take advantage of information from the previous inputs/outputs, to enhance its predictive capability.

Time series analysis can be applied to the output sequence produced by a DNN to identify the trend and predict future values. Among the numerous models available for time series analysis, the most widely used ones are autoregressive (AR), integrated (I) and moving average (MA) models, along with their combinations. An AR model of order $k$ predicts the next value $x_t$ as a linear combination of past values $x_{t-1}, \ldots, x_{t-k}$:

\begin{equation} \label{eq:ar}
x_t = \alpha_0 + \sum_{i = 1}^k{\alpha_i x_{t-i}} + \epsilon_t
\end{equation}

\noindent
where coefficients $\alpha_0, \ldots, \alpha_k$ can be estimated by the least square method and $\epsilon_t$ represents the error term.

Processing of a sequence of inputs can be achieved by recurrent neural networks (RNN) equipped with long short-term memory (LSTM)~\cite{Hochreiter:1997:LSM:1246443.1246450}, which is capable of dealing with both short and long range dependencies. 
In LSTM, outputs are influenced not only by the current input but also by the state of the RNN, which encodes the entire history of past inputs. A \textit{forget gate} layer of LSTM decides what information to keep/remove from the previous network state whereas an \textit{additive gate} decides how to update the state based on the current input. 

\section{Problem formulation}\label{sec:problem}


We focus on SDCs that perform \textit{behavioural cloning}, i.e., the DNN learns the \textit{lane keeping}~\cite{Gambi:2019:ATS:3293882.3330566} behaviour from a human driver.
Models such as the ones by NVIDIA~\cite{nvidia-dave2} or the Udacity self-driving challenge~\cite{udacity-challenge} are trained with visual inputs (i.e., images) from car-mounted cameras that record the driving scene, paired with the steering angles from the human driver. The DNN then ``learns how to drive'' by discovering underlying patterns within the training images representing the shape of the road, and predicting the corresponding steering angle. 


\begin{figure*}[t]
\centering
\includegraphics[trim=0cm 23cm 0cm 0cm, clip=true,scale=0.20]{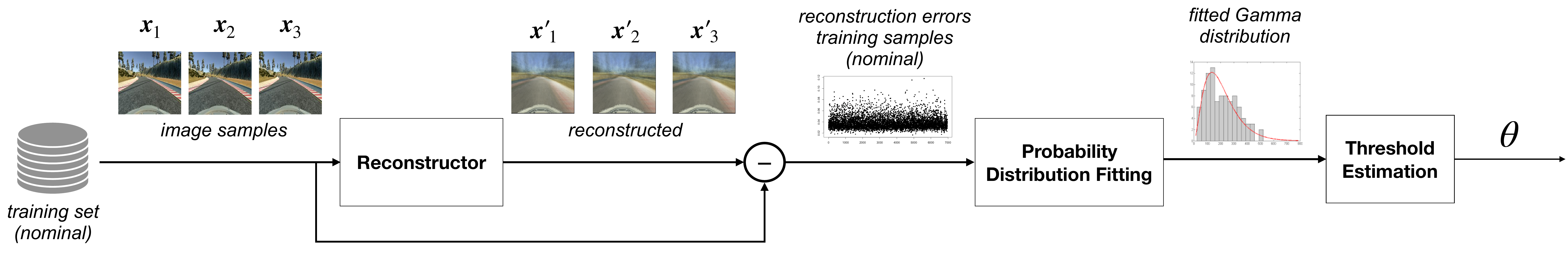}
\caption{Model Training under Nominal Driving Behaviour.} 
\label{fig:approach} 
\end{figure*}

For a classification problem (e.g., hand-written digit recognition), a misbehaviour can be defined as an input that can be confidently labeled by humans while it is misclassified by a DNN. 
Differently, for a regression problem 
such a definition is more troublesome, because there is no expected outcome for an individual output of the DNN and it is only the overall behaviour resulting from the DNN predictions that may or may not be acceptable, depending on the specific application domain.

In steering angle prediction, it is challenging to decide if the steering angle produced by a DNN is wrong, because the optimal steering angle is generally unknown for a new test scenario and even if it were known, the amount of difference between predicted and expected steering angle that qualifies as an error is difficult to decide a priori. It is instead more realistic to figure out whether a chain of inaccurate predictions, regardless of the amount of deviation from the optimal angles, ultimately leads to a misbehaviour of the system. 

In fact, the  definition of misbehaviour should be decoupled from the notion of correct/wrong DNN output, being instead linked to the ability of a DNN of abstracting from the training examples and learning how to drive in different ways/conditions. 
It is the task of the DNN to  generalize the training knowledge to make the model robust w.r.t. slightly different conditions from those observed in the training set.
For instance, if the SDC is facing a 90-degree bend, the DNN may decide to steer at a certain steering angle $\theta$ if the car speed is 60 mph, or $\frac{\theta}{2}$ if the speed is 30 mph.

\begin{defn}[\textbf{Misbehaviour of a DNN}]
A DNN exhibits a \textit{misbehaviour} in a given test scenario if the overall system that contains the DNN does not respect its requirements due to the outputs produced by the DNN.
\end{defn}

In the autonomous driving domain, there are many possible misbehaviours, associated with the different requirements that such systems are supposed to realize. Safety requirement violations are by far the most critical requirements, as a misbehaviour in the steering component may cause a crash of the vehicle with potential casualties.
However, in general, a SDC might violate also other driving requirements, e.g., related to ride comfort~\cite{10.1007/978-3-658-21194-3_53}, such as excessive derivative of the steering angle, unstable movement around the centerline, or excessive deceleration. All these could also be considered as misbehaviours by our definition.

In this paper, we focus on the prediction of two safety-critical misbehaviours: (1)~collisions, and (2)~out-of-bound episodes (OBEs). 
The rationale for this choice are as follows. 
First, they represent the vital requirement to be satisfied and thoroughly tested (i.e., the car should stay in lane and avoid whatsoever collision), without which autonomous driving vehicles would be hardly accepted in production.
Moreover, leveraging a simulation environment such as Udacity's~\cite{udacity-simulator}, allows us to: (1)~safely test such critical scenarios, (2)~precisely define, observe and measure them, in order to support crash analysis and reproduction.

Thus, the \textbf{\textit{problems}} we want to address in this paper are: (1)~recognizing when a self-driving car is within a low-confidence area because of an unexpected execution context, and (2)~predicting such situation timely enough so as to take countermeasures before the vehicle may crash or drive off road.




\section{Approach}\label{sec:approach}


The goal of our approach is to monitor the confidence level of a SDC as it runs and to promptly predict whether drops in confidence correlate with potential future misbehaviours. 
Our approach works in an end-to-end fashion, analyzing directly the input data as retrieved by the car (an image from the center camera), making the approach independent from the specific architecture of the self-driving component, requiring no modifications to the existing DNN model, and being therefore, highly generalizable.

The main working assumption is that a prediction model trained on normal data should learn the normal time series patterns. When the model is used on a SDC in the field, it should worsen its performance as the car approaches previously unseen regions as compared to normal, known regions (\autoref{fig:example}). Then, what's needed is a way to set up a good decision boundary in order to timely alert the human driver (NHTSA Level~4) or the main self-driving component (NHTSA Level~5) about triggering self-healing. 

We now detail each step of our approach, which consists of two main phases: (1)~\textit{model training under nominal driving behaviour}, and (2)~\textit{field usage of the trained model}.

\subsection{Training of \tool under Nominal Driving Behaviour}\label{sec:training}

\autoref{fig:approach} illustrates the training phase of our approach, which consists of several steps. 

\subsubsection{Reconstructor}

The first step consists in retrieving a model of normality from the training driving scenarios. Thus, in the training set, we capture the visual input stream of the SDC under nominal situations.


Then, we train our driving scenario reconstructor with such ``normal'' instances.
Let us consider a training set $\boldsymbol{X} = \{x_1, x_2, \ldots, x_n \}$ of $n$ image frames, where the index $i \in [1:n]$ of $x_i \in \boldsymbol{X}$ represents the discrete time $t$. 

Depending on the considered architecture, a reconstructor can be \textit{singled-image} or \textit{sequence-based}. 
For singled-image reconstructors, only one image frame is considered at a time. When the discrete time is $t = i$, $x_i$ is the input and the reconstructor recreates it into $x'_i$. 
For sequence-based reconstructors, assuming $k$ image frames preceding $x_i$ are used to reconstruct $x_i$, the sequence 
$\langle x_{i-k}, \ldots, x_{i-1}\rangle$ is the input used to output $x'_i$, a prediction of the actual current frame $x_i$.
For instance, for $k=3$ and $i = 4$, the reconstructor considers the sequence $\langle x_1, x_2, x_3 \rangle$ in order to predict the current frame $x_4$.

At the end of this task, each reconstruction error $e_i=d(x_i, x'_i)$ can be computed, where $d$ is a proper distance function (e.g., Euclidean distance). This results in the set of reconstruction errors $\boldsymbol{E} = \{e_1, e_2, \ldots, e_n \}$, available for all elements in the training set $\boldsymbol{X} = \{x_1, x_2, \ldots, x_n \}$.

\begin{figure}[b]
\centering
\fbox{
\includegraphics[trim=0cm 15cm 11cm 0cm, clip=true,scale=0.35]{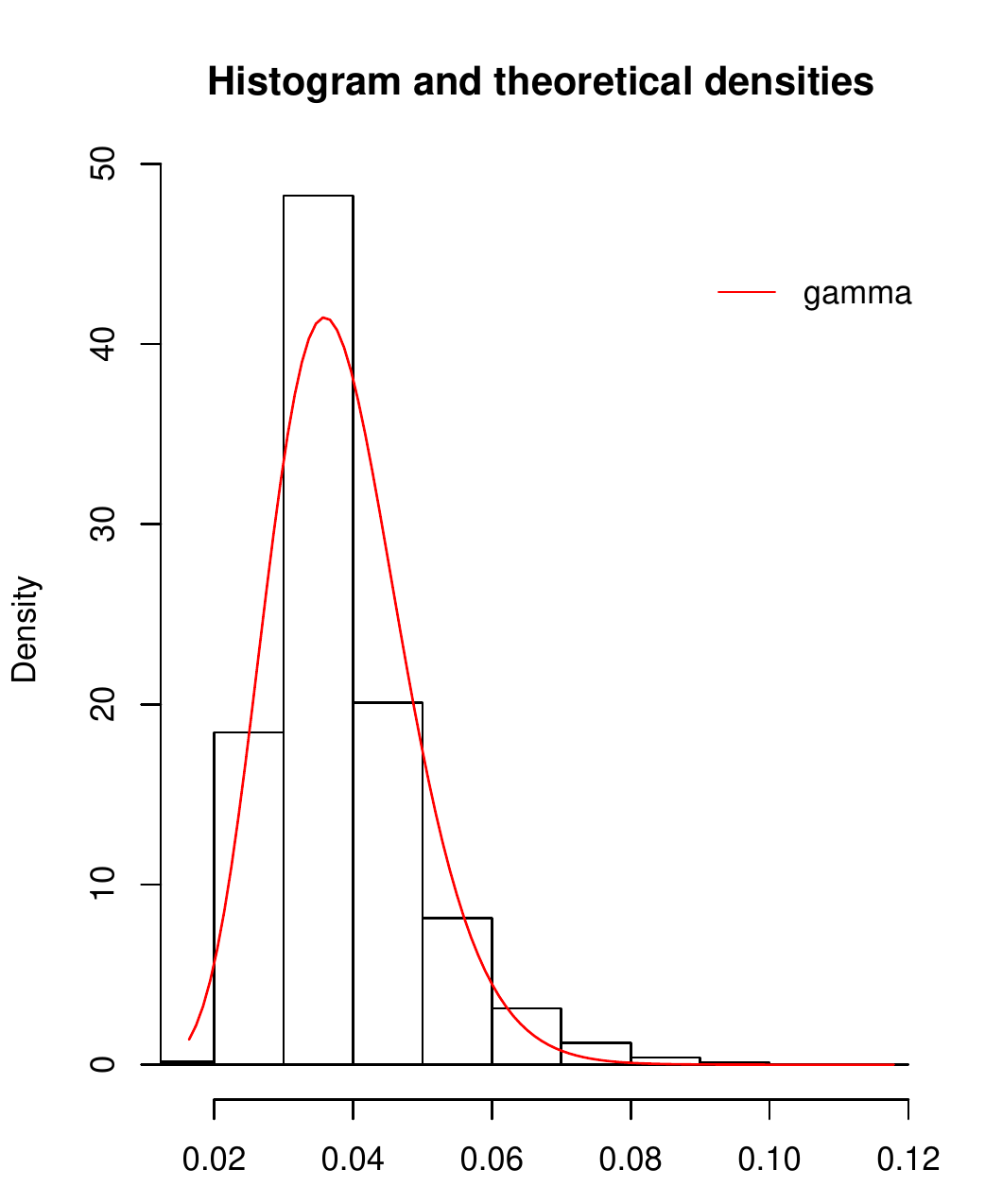}
}
\caption{Fitted Gamma distribution of reconstruction errors from a VAE on the Dave2 dataset.} 
\label{fig:hist} 
\end{figure}

\subsubsection{Probability Distribution Fitting}

We build a model of normality for the reconstruction errors $\boldsymbol{E} = \{e_1, e_2, \ldots, e_n \}$ collected in nominal driving conditions.
We use probability distribution fitting to obtain a statistical model of normality and using such model we determine a threshold $\theta$ that brings the expected false alarm rate in nominal conditions below some acceptable, configurable level $\epsilon$ (e.g., $\epsilon = 10^{-3}$ or $\epsilon = 10^{-4}$).


The reconstruction error $e = d(x, x')$ can be computed by comparing the individual pixels of the images $x$ and $x'$ and taking the mean pixel-wise squared error. Assuming images have width $W$, height $H$ and $C$ channels (usually, 3 RGB channels for colour images), the reconstruction error is defined as follows:

\begin{equation}
d(x, x') = \frac{1}{W H C} \sum_{i = 1, j = 1, c = 1}^{W, H, C}{(x[c][i, j] - x'[c][i,j])^2}
\end{equation}

With some reasonable approximation, we can assume that the pixel-wise error $e[c][i, j] = x[c][i, j] - x'[c][i,j]$ follows a normal distribution with pixel dependent variance: $e[c][i, j] \sim \mathcal{N}(0, \sigma_{c, i, j})$. Correspondingly, the sum of the squares of pixel-wise errors $e[c][i, j]$ will be a gamma distribution: $e = d(x, x') \sim \Gamma(\alpha, \beta)$. We get a gamma instead of a $\chi^2$ distribution because pixel-wise errors have different (channel/pixel dependent) and non-unitary variances.

%
%
%

\head{Definition of  Gamma Distribution}
Gamma is a probability model for a continuous variable on $[0,\infty)$ which is widely used in engineering, science, and business, to model continuous variables that are always positive and have skewed distributions. 

The \textit{probability density function} of a random variable $x\sim\Gamma(\alpha, \beta)$ is:

\begin{equation}\label{eq:pdf-gamma}
f(x)=\frac{\beta^{\alpha}}{\Gamma(\alpha)}x^{\alpha-1}e^{-\beta x} 
\hspace{.2in} x > 0; \ \alpha, \beta > 0
\end{equation}

\noindent
where $\alpha$ is the \textit{shape} parameter (which affects the shape of the distribution), $\beta$ is the \textit{rate} parameter (or inverse scale, which stretches/shrinks the distribution) and $\Gamma$ is the gamma function. 
When $\alpha$ is large, the gamma distribution closely approximates a normal distribution with the advantage that the gamma distribution has non-zero density only for positive real numbers.

The \textit{gamma function} $\Gamma$ can be seen as a solution to the interpolation problem of finding a smooth curve that connects the points  $(n, m)$ with $m = (n-1)!$ at any positive integer value for $n$. 
Such a definition was extended to all complex numbers with a positive real part by Bernoulli, as a solution to the following integral:

\begin{equation}\label{eq:gamma}
\Gamma(z) = \int_{0}^{\infty} {x^{z-1}e^{-x}dx} 
\hspace{.2in}  \mathfrak{R}(z) > 0;
\end{equation}

\begin{figure*}[t]
\centering
\includegraphics[trim=0cm 27.5cm 3cm 0cm, clip=true,scale=0.205]{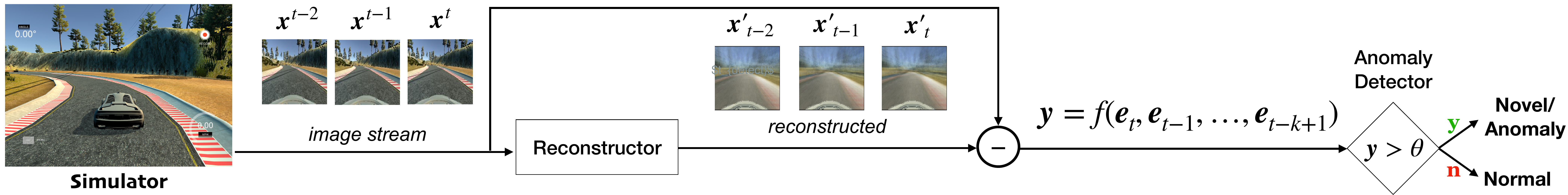}
\caption{Usage Scenario of \tool. 
} 
\label{fig:testing} 
\end{figure*}

\head{Fitting the Gamma Distribution}
One powerful method to estimate the parameters of a (gamma) distribution that fit the data (reconstruction errors) best is by maximum likelihood estimation (MLE). The likelihood function reverses the roles of the variables: in \autoref{eq:pdf-gamma}, the values of $x$ are known, so they are the fixed constants, whereas the unknown variables are the parameters $\alpha$ and $\beta$. MLE involves calculating the values of these parameters so as to obtain the highest likelihood of observing the values of $x$ when the given parameters are supplied to $f$.

Under the assumption of independence of the data, the likelihood of the data given the parameters of the distribution is conveniently defined as the logarithm of the joint probability of the data for a given choice of the parameters. In the case of a gamma distribution, with a dataset consisting of reconstruction errors $\boldsymbol{E} = \{e_1, e_2, \ldots, e_n \}$, we get:

\begin{eqnarray} \label{eq:maxlike}
\mathcal{L}(\alpha, \beta; \boldsymbol{E}) = \frac{1}{n}\sum_{i=1}^n{\log f(e_i | \alpha, \beta)} = \\
  n (\alpha-1)\overline{\log e} - n \log \Gamma(\alpha) - n \alpha \log \beta - n \overline{e} / \beta
\end{eqnarray}

\noindent
where $\overline{e}$ ($\overline{\log e}$) is the mean (log) reconstruction error over $\boldsymbol{E}$.
To find the values of parameters $\alpha$ and $\beta$ that maximize $\mathcal{L}$ we have to find a solution to the equations: 
(1) $\partial \mathcal{L} / \partial \alpha = 0$; (2) $\partial \mathcal{L} / \partial \beta = 0$.
%
%
The second equation can be easily solved analytically, resulting in $\beta = \overline{e} / \alpha$. By substituting the value of $\beta$ into the first equation we get an equation that unfortunately cannot be solved analytically. However, the Newton method can iteratively converge to the solution quite quickly. The output of such numerical estimation will be the pair of parameters $\alpha$ and $\beta$ of the gamma distribution that best fit the reconstruction errors.

\head{Example of Threshold Estimation}
Let us consider a set of reconstruction errors. \autoref{fig:hist} shows the histogram of those produced on the Dave2 dataset when the reconstructor is a VAE. On such dataset the MLE method estimates the following gamma parameters: $\alpha = 15$; $\beta = 392$. With these parameter values, the plot of the gamma distribution is the red line in \autoref{fig:hist}. We can notice a good agreement between the histogram and the estimated probability distribution.

Let us assume that we are willing to accept a false alarm rate $\epsilon = 10^{-2}$. The threshold $\theta$ with a probability mass above the threshold equal to $10^{-2}$ can be easily obtained as the inverse of the cumulative gamma distribution $F(x)$: $\theta = F^{-1}(1-\epsilon)$. This ensures that the cumulative probability of values $\leq \theta$ is $1-\epsilon$, leaving only a probability of $\epsilon$ to the tail following $\theta$. 
We use the estimated $\theta$ as threshold to distinguish anomalous conditions (reconstruction error $\geq \theta$) from normal ones (reconstruction error $< \theta$).

\subsection{Usage Scenario}\label{sec:usage-scenario}

\autoref{fig:testing} shows how \tool is used online for misbehaviour prediction after model training (i.e., after fitting the gamma distribution and estimating the threshold $\theta$). Misbehaviour prediction is executed online as the SDC drives. 
In this phase, the SDC generates data continuously and the reconstructor recreates the incoming stream of images. The sequence of reconstruction errors is passed through the autoregressive model $f$ and the resulting, filtered error is compared against the threshold $\theta$, which determines whether an anomaly is detected or not. In the former case, self-healing is triggered and the SDC is brought to a safe state.

\subsubsection{Time-aware Anomaly Score Prediction}\label{sec:time-aware-anomaly-score}

The reconstruction error $e_t$ at time $t$ might be susceptible to single-frame outliers, which are not expected to have a big impact on the driving of the car, but would indeed make the misbehaviour predictor falsely report an anomalous context.
For this reason, we smooth such noisy oscillations by applying an autoregressive filter:
the sequence of reconstruction errors is passed to a module that performs time series-analysis. In \autoref{fig:testing}, this corresponds to the AR filter $f$. The output of this filter, instead of the raw reconstruction error $e_t$, is compared with the threshold $\theta$ to recognize unexpected driving conditions. In our experiments we used a simple AR model (see \autoref{eq:ar}) with $\alpha_0 = 0$ and $\alpha_i = 1/k$ for $i = 1, \ldots, k$.

%
%
%

%


\section{Empirical Evaluation}\label{sec:evaluation}

\subsection{Research Questions}

We consider the following research questions:

\noindent
\textbf{RQ\textsubscript{1} (effectiveness):}
How effective is \tool in predicting anomalies for autonomous vehicles? What are the best reconstructors to use?

\noindent
\textbf{RQ\textsubscript{2} (prediction):}
How does the misbehaviour predictions of \tool change as we increase the reaction period (i.e., we anticipate the time of prediction)?

\noindent
\textbf{RQ\textsubscript{3} (comparison):}
How does \tool compare with DeepRoad's \cite{deeproad} online input validation?

\subsection{Self Driving Car Models}
We evaluate our framework on three existing DNN-based SDCs: NVIDIA's DAVE-2~\cite{nvidia-dave2}, Epoch~\cite{epoch}, and Chauffeur~\cite{chauffeur}.
We choose these models because they are publicly available, thus they can be trained and evaluated on the simulator. Moreover, they have been objects of study of other testing works~\cite{deepxplore,deeptest}.
DAVE-2 consists of three CNNs, followed by five fully-connected layers. Chauffeur uses a CNN to extract the features of input images, and a RNN to predict the steering angle from 100 previous consecutive images. Epoch consists of a single CNN model.

\subsection{Simulation Platform}\label{sec:simulator}

A major problem in anomaly detection research is the lack of labeled benchmark datasets~\cite{Campos:2016:EUO:2962863.2962870}, and the self-driving car domain is no exception.
Unlike previous works~\cite{deeproad,deeptest,deepxplore}, we cannot rely on existing driving image datasets such as the ones released by Udacity~\cite{udacity-datasets}, because they lack any episode of crash, or cars driving off road whatsoever. Moreover, our definition of misbehaviour (\autoref{sec:problem}) requires the creation of a set of ``controllable'' unexpected conditions that may potentially cause them, along with a way to precisely record them.
Thus, to investigate the effectiveness of our approach in predicting safety-critical misbehaviours, we evaluated \tool in the Udacity simulator~\cite{udacity-simulator}.

The Udacity simulator is developed with Unity~\cite{unity}, a popular cross-platform game engine.
The simulator provides two default tracks, 
for testing DNNs models. The simulator executes in two modes: (1)~\textit{training mode}, in which the user manually controls the car while the simulator records her actions, and (2)~\textit{autonomous mode}, in which the car is controlled by an external agent, such as a DNN-based autonomous driving system.
Moreover, we added a third track~\cite{mountain-track} to the existing set,
%
%
%
%
and we implemented two additional components, namely, an \textit{unexpected context generator}, and a \textit{collision/OBE detection system}.

\subsubsection{Unexpected Context Generator}\label{sec:ucg}

\begin{figure}[H]
    \centering
    \includegraphics[width=0.3\linewidth]{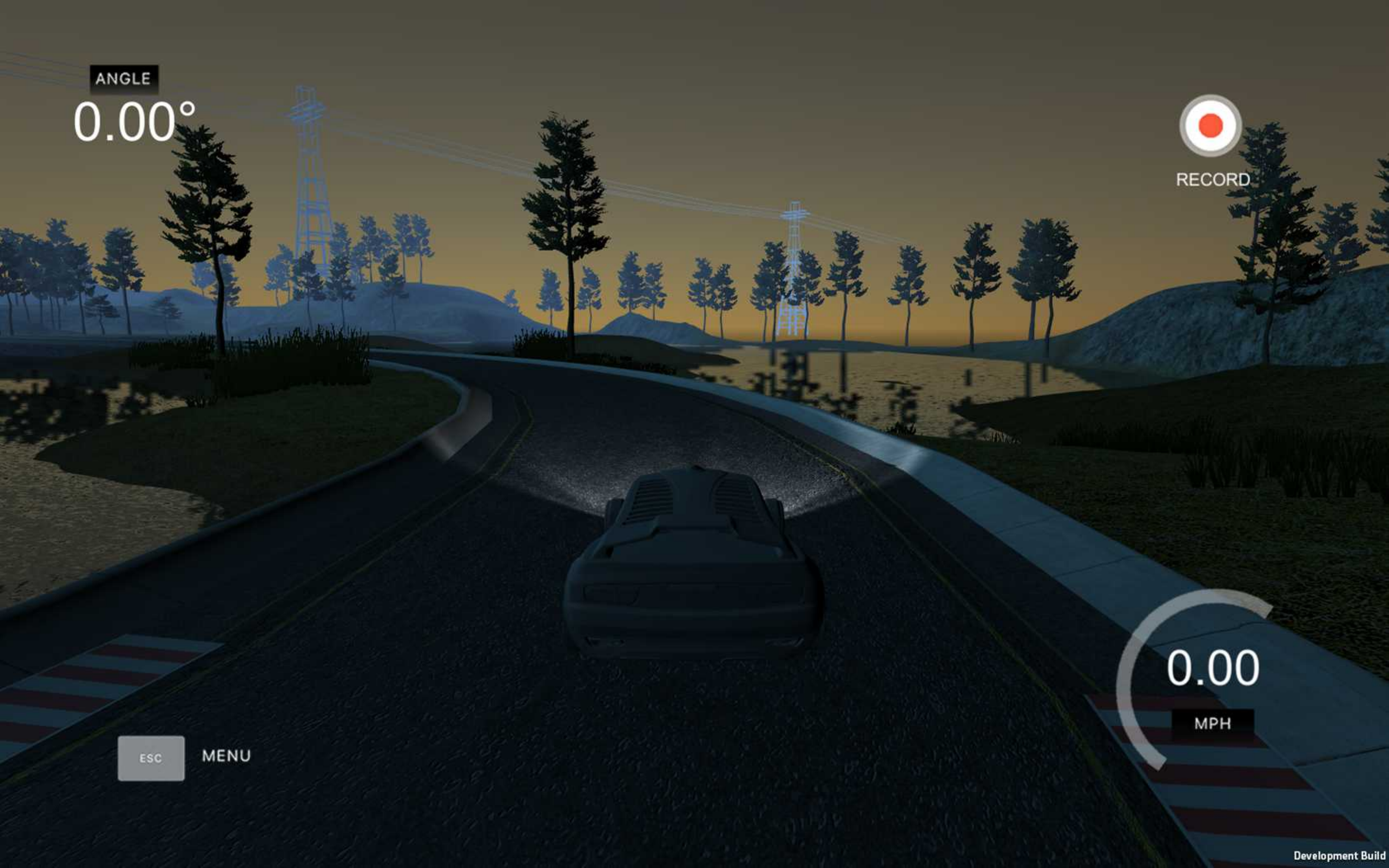}
    \includegraphics[width=0.3\linewidth]{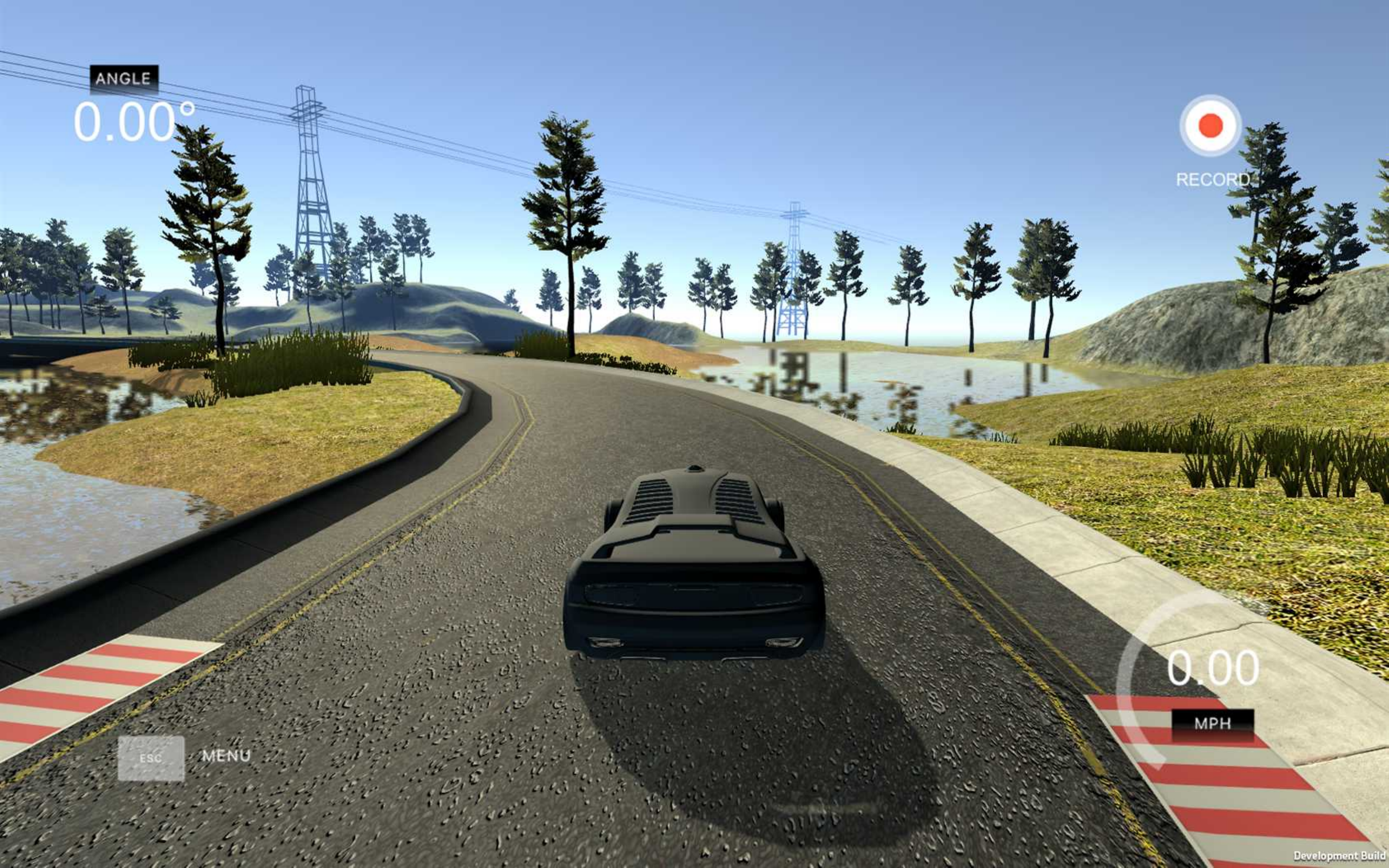}
    \includegraphics[width=0.3\linewidth]{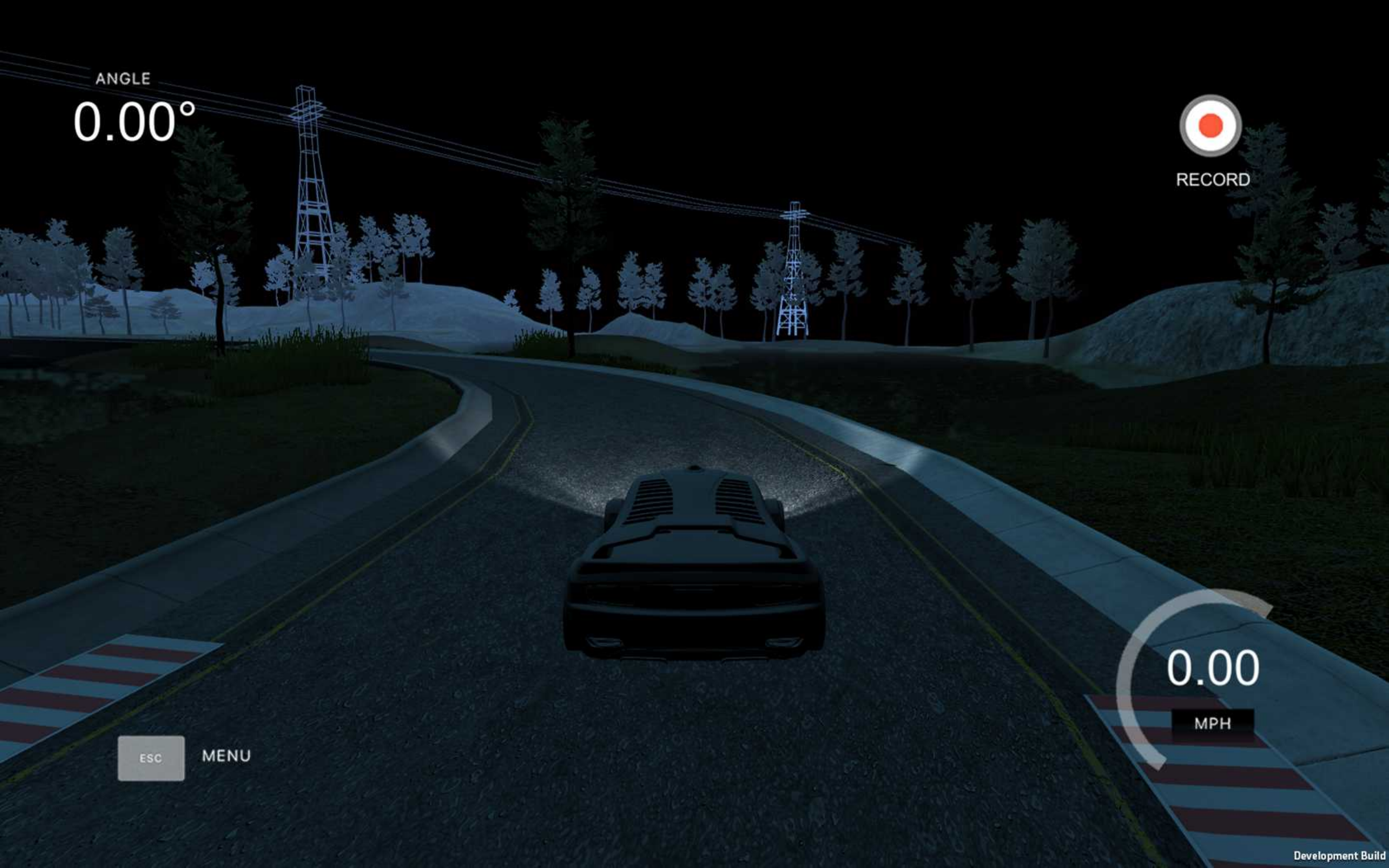}
    \includegraphics[width=0.3\linewidth]{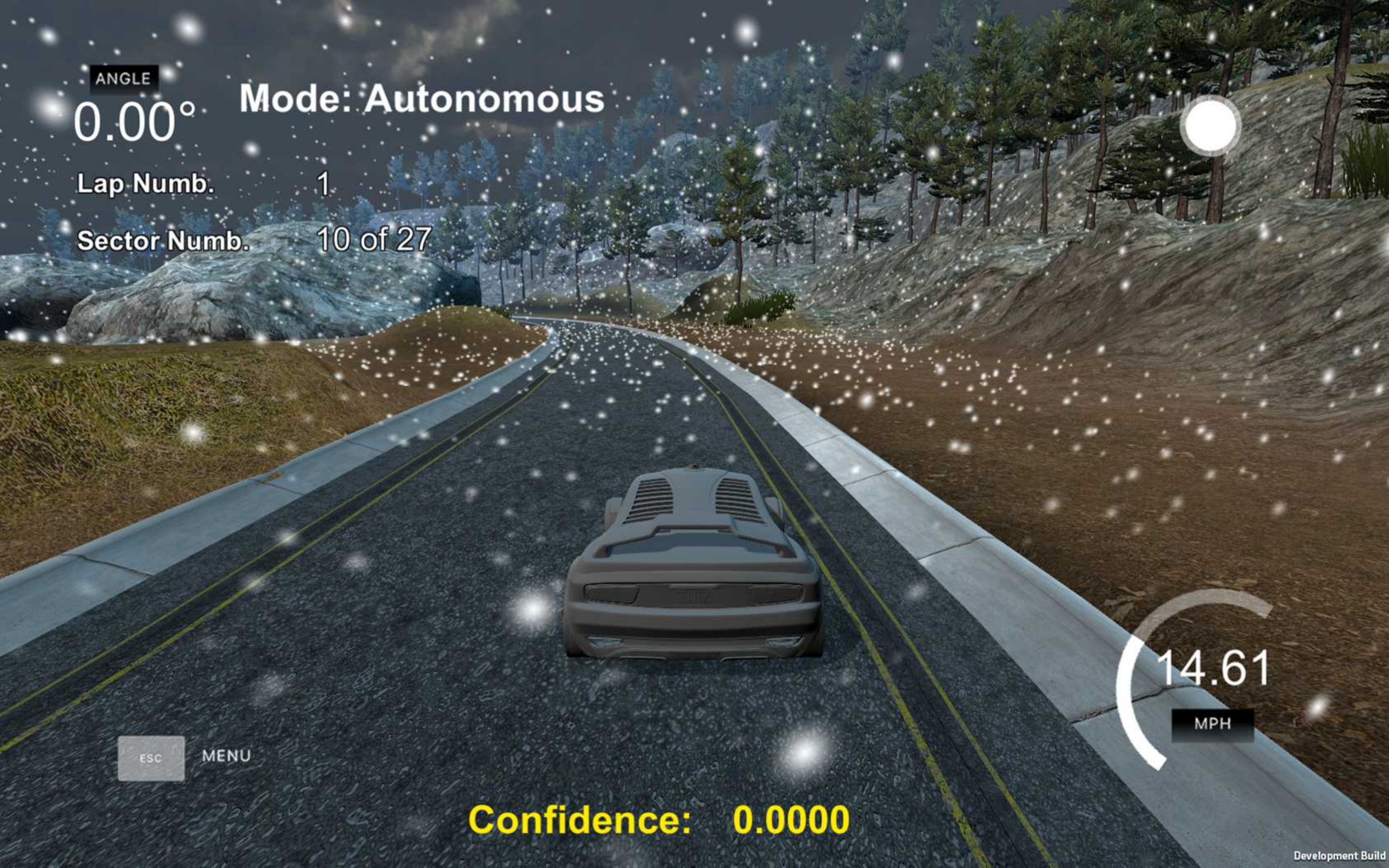}
    \includegraphics[width=0.3\linewidth]{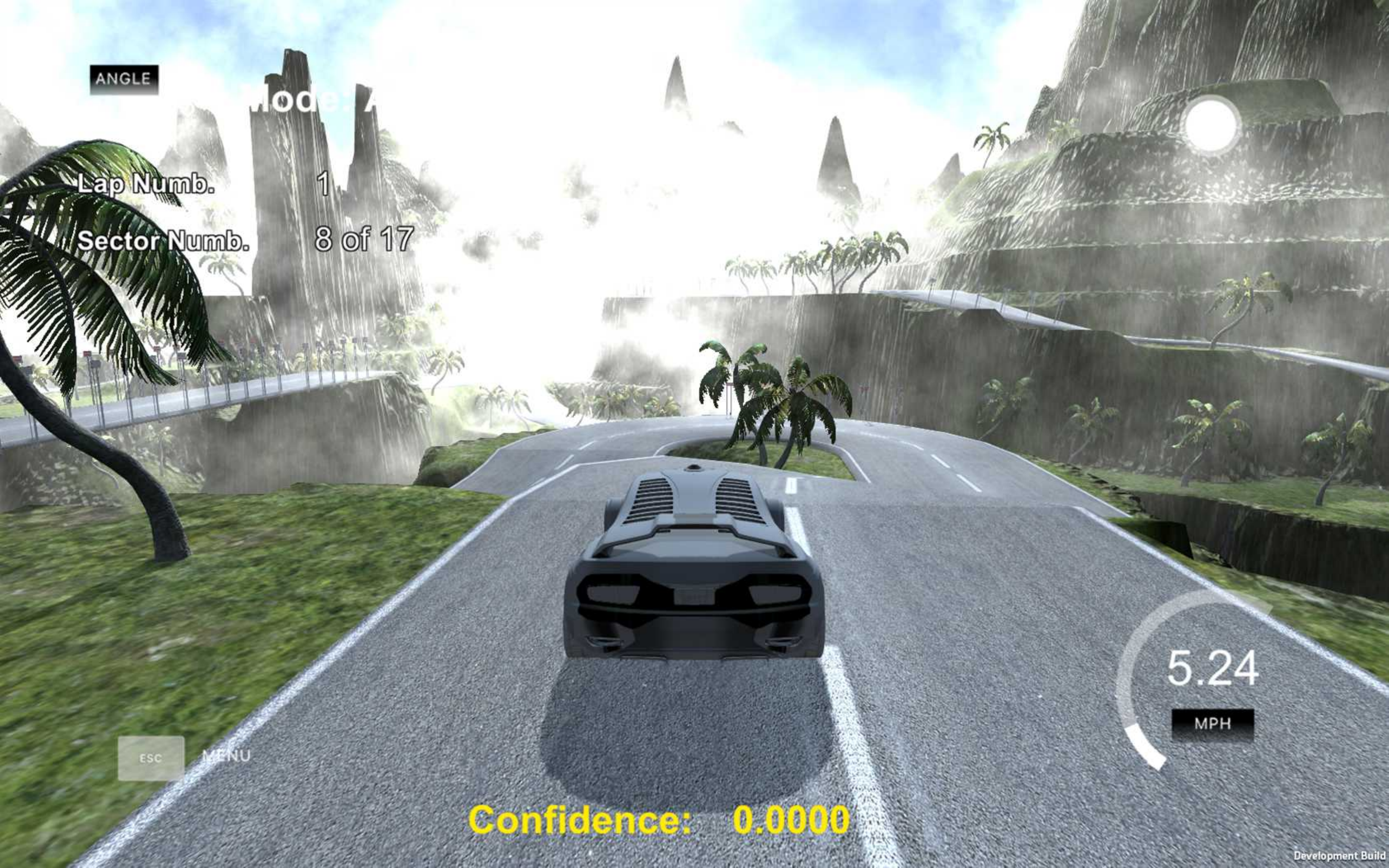}
    \includegraphics[width=0.3\linewidth]{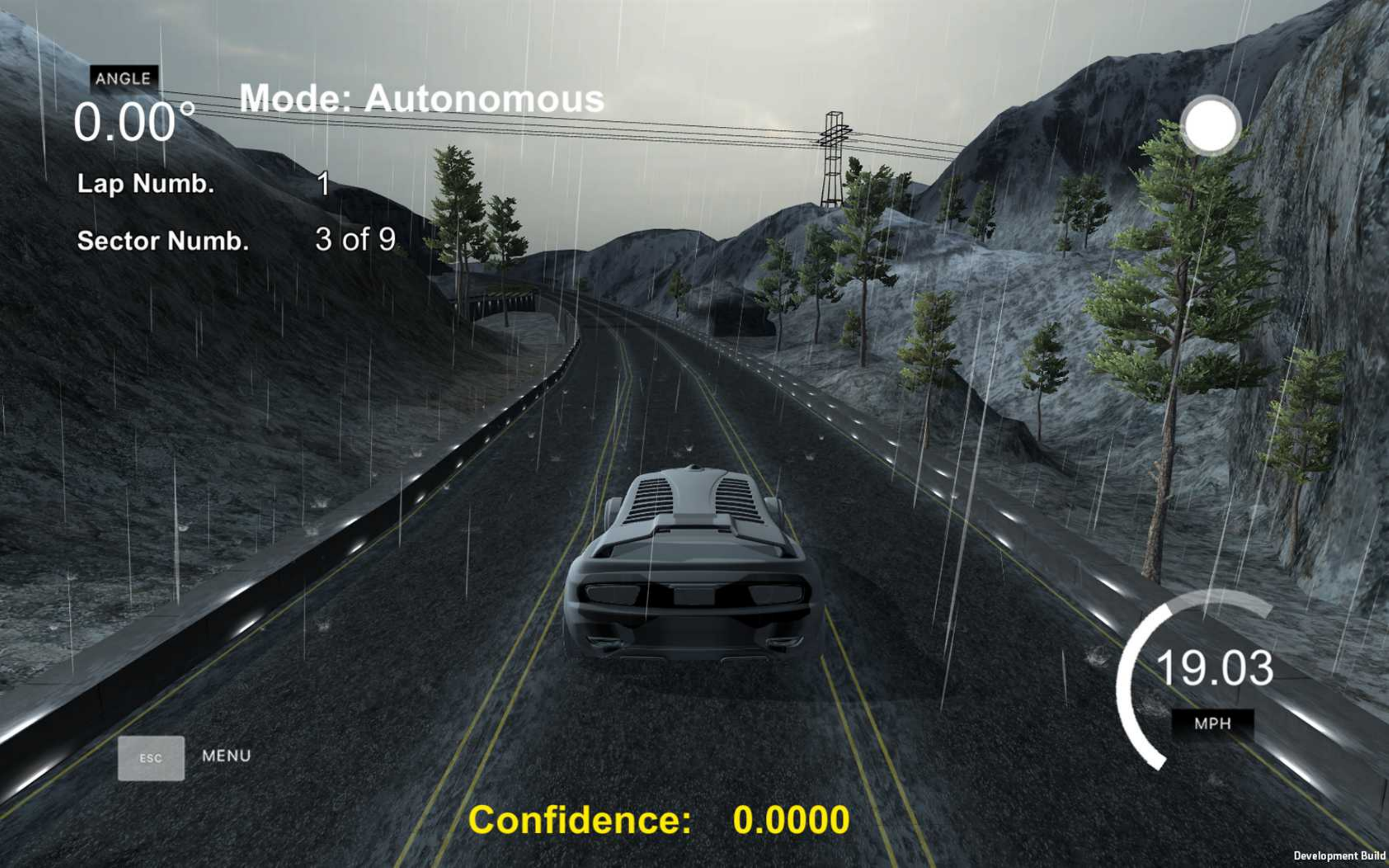}
    \caption{(top) Day/night cycle (sunrise, day, night) and (bottom) weather effects (snowy Lake Track, foggy Jungle Track, and rainy Mountain Track).}
    \label{fig:effects}
\end{figure}

First, we developed a method to gradually inject unseen conditions during testing mode (i.e., conditions diverse from the training mode's defaults).
The first condition deals with illumination and introduces a \textit{day/night cycle} component to gradually change the light condition of the track during the simulation. The effect is customizable and consists of smooth increase/decrease of brightness/darkness over a fixed period (60~s in our experiments).
The second kind of unexpected condition deals with \textit{weather}. We implemented rain, snow and mist effects, with a variable intensity during the simulation. For the implementation, we used a specific Unity component called Particle System~\cite{particle-system}
which can simulate the physics of a cluster of particles with high performance. Rain particles emission rate ranges between a minimum of 100 (light rain) to a maximum of 10,000 particles/s (heavy rain); fog between [100..2,000] particles/s, and snow between [100..800] particles/s.
\autoref{fig:effects} shows a few examples.

\subsubsection{Collision/OBE Detection System}

Following our definition of safety-critical misbehaviours, we implemented an automated collision/OBE detection system (ACODS) that records any unwanted interaction of the SDC with the environment, allowing us to experiment the effectiveness of \tool at anticipating such episodes during the occurrence of unexpected scenarios (\autoref{sec:ucg}).

\begin{figure}[ht]
    \centering
    \includegraphics[width=0.4\linewidth]{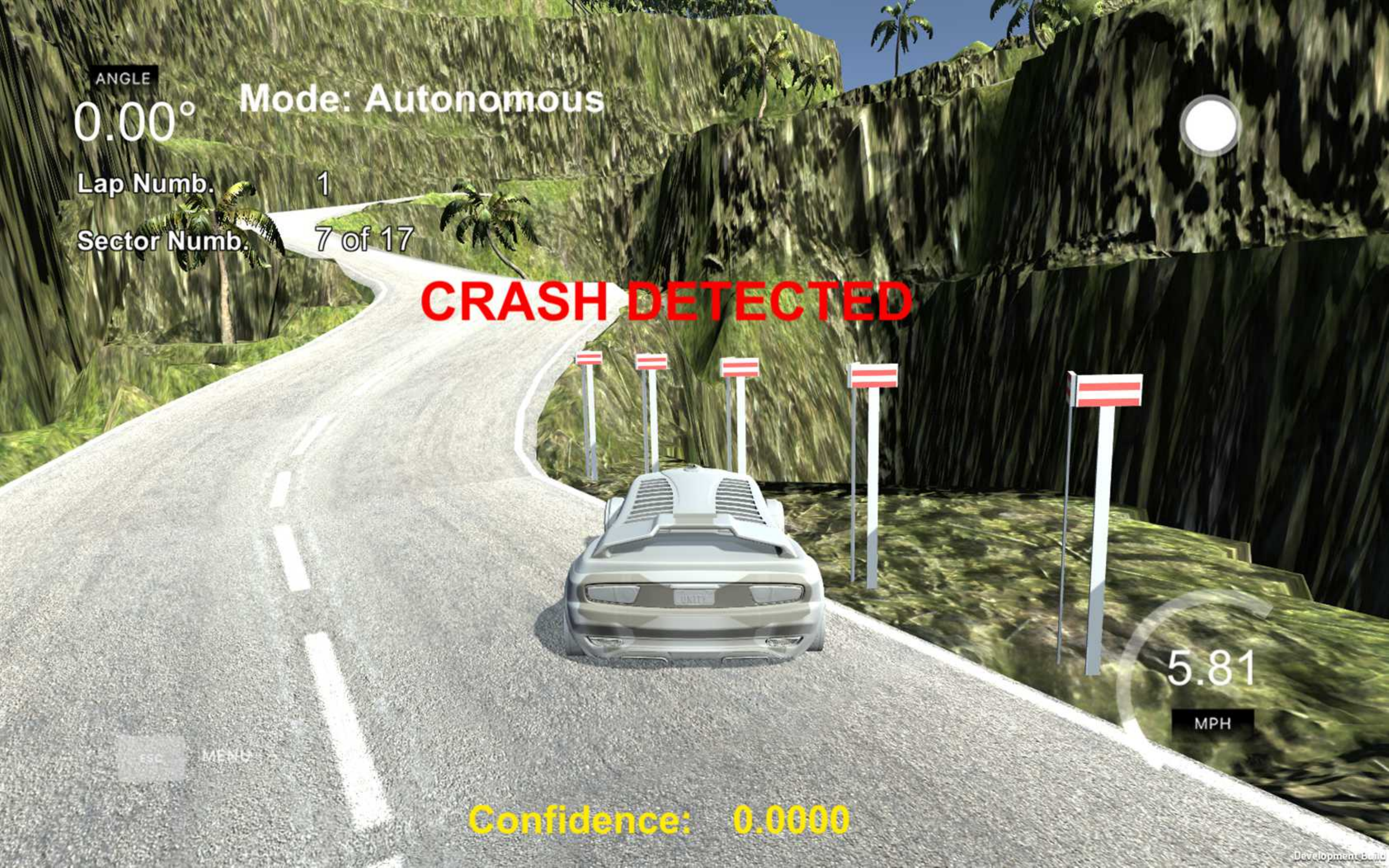}
    \includegraphics[width=0.4\linewidth]{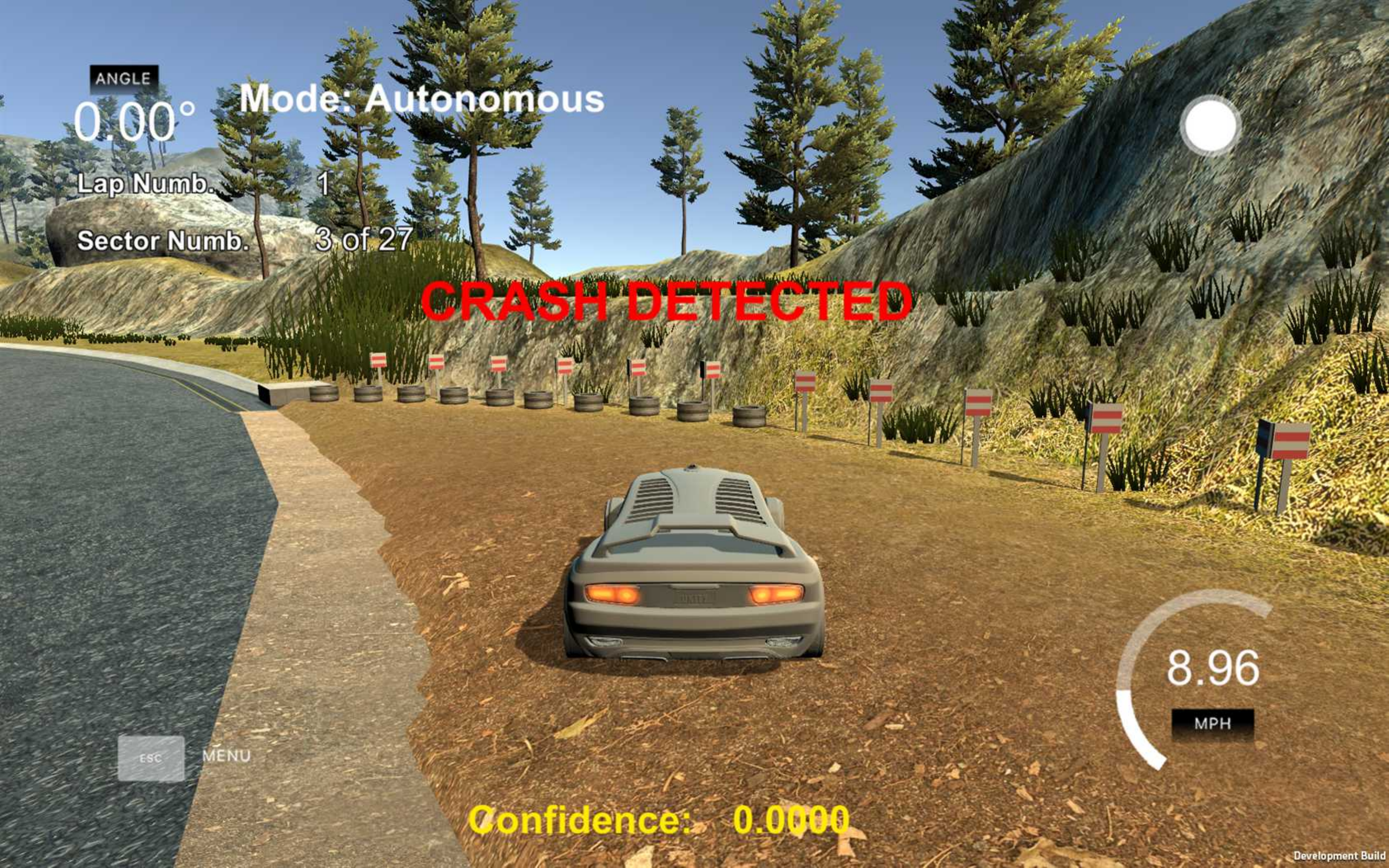}
    \caption{Simulator crash/OBE detection.}
    \label{fig:crash-detection}
\end{figure}

We implemented ACODS based on \textit{colliders}, which are consolidated building blocks of modern game engines to simulate the physical interaction between objects. We approximate the car body with a geometry mesh, and implemented a collider callback that informs the simulator of any physical interaction of the car with scene objects. When the car ``hits'' the road, then it means the car is actually on track, whereas when the car ``collides'' with any other object, the callback registers whether it is a crash against some object (\autoref{fig:crash-detection}~(left)), or whether it is an OBE (\autoref{fig:crash-detection}~(right)).

Secondly, we implemented an automatic restart mechanism that restores the SDC to a safe position after a crash/OBE, allowing us to record multiple simulations without the need for manual restart.

\subsection{Procedure}

\subsubsection{Data Generation (Training Set)}\label{sec:data_generation}

Training data were collected by the authors in training mode by performing 10 laps on each track, following two different track orientations (normal, reverse). Overall, we obtained a dataset of 124,638 training images (at 10-13 fps), divided as follows: 32,243 for Track 1 (Lake), 51,422 for Track 2 (Jungle), and 40,973 for Track 3 (Mountain). Differences depend on the track lengths.
To allow a smooth driving and a correct behaviour capture (i.e., lane keeping), the maximum driving speed was set to 30 mph, the default in the Udacity simulator.

\subsubsection{SDC Model Setup \& Training}\label{sec:model-training}

All SDCs models were trained on 41,546 images from the central camera.
We used data augmentation as a consolidated practice for building more reliable and generalizable SDCs, limiting the lack of image diversity in the training data. Specifically, 60\% of the data was augmented through different image transformation techniques (e.g., flipping, translation, shadowing, brightness). We cropped the images to 66 x 200, and converted them from RGB to YUV colour space.
All SDC models were trained for 500 epochs with batch size of 256 on a machine featuring an i9 processor, 32 GB of memory, and an Nvidia GPU 2080 TI with 11GB of memory. Basically, the training was meant to create solid models for testing, i.e., able to drive multiple laps on each track under nominal conditions without showing any misbehaviour in terms of crash/OBE.

\subsubsection{Evaluation Set}

To collect the evaluation data, we executed 72 simulations (2 laps each) in autonomous mode (3 SDC x 8 conditions x 3 tracks). As in data generation, the maximum speed was set to 30 mph. Specifically, for each SDC and for each track, we performed 1 simulation in the same normal conditions as the training set. This allows us to estimate the number of false alarms (false positives) in nominal conditions. 
Second, we performed 4 simulations activating in turn a single unexpected condition: day/night cycle, rain, snow, fog. Third, we performed 3 simulations activating in turn a combined condition: day/night cycle + rain, day/night cycle + snow, day/night cycle + fog.


In our experiments, we used a value of 60~s both for the day/night cycle and for the loop between the minimum and maximum intensity of the effects. This value was chosen empirically given the relatively short speed of the SDC, and the small length of the tracks, and allowed us to test the behaviour of the SDC on each of the track subsets under all possible conditions. For example, the bridge part of Track1 (Lake) has been driven on under both dawn/day/sunset/night conditions (day/night cycle) and minimal/maximal intensity of rain/fog/snow.

Overall, we obtained a dataset of 778,592 images, divided as follows: 188,032 for Track 1 (Lake), 260,064 for Track 2 (Jungle), and 208,064 for Track 3 (Mountain) with unknown conditions and 33,088 for Track 1, 46,272 for Track 2 and 43,072 for Track 3 in known conditions.

\begin{figure*}[t]
    \centering
    \includegraphics[scale=0.8]{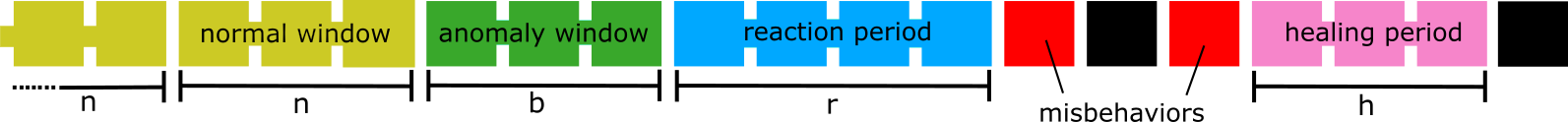}
    \caption{Labelling of Anomalous and Normal Windows in Driving Stream.}
    \label{fig:labelling}
\end{figure*}

\subsubsection{\tool's Configurations}

We used four autoencoders taken from existing guidelines~\cite{keras-autoencoders}: 
(1)~\emph{SAE} (simple autoencoder with a single hidden layer),
(2)~\emph{DAE} (deep five layers fully-connected autoencoder),
(3)~\emph{CAE} (convolutional autoencoder alternating convolutional and max-pooling layers), and
(4)~\emph{VAE (variational autoencoder)}.

All autoencoders take as input a single image. We added images taken by the side cameras of the car (left, right) to allow better generalization, even though, during testing, autoencoders are not used for prediction. Lastly, we performed further data augmentation on 60\% of the inputs, as described in \autoref{sec:model-training}.

As an additional sequence-based reconstructor, we also implemented an LSTM consisting of two LSTM-layers and one convolutional layer.
%
For implementation details, we made our code publicly available in the replication package accompanying this paper~\cite{tool}.

\head{Baseline}
We use the input validation technique of \deeproad~\cite{deeproad} as baseline for \tool. 
Unfortunately, authors did not make their code available; therefore, we implemented of our own version according to the description in the paper.
For input validation, \deeproad uses the pre-trained VGG19 ImageNet classifier~\cite{Simonyan2014} to extract style and feature vectors from a given image. Principal Component Analysis (PCA) is then used to reduce all style and feature vectors, concatenated into a matrix, to three dimensional representations, which support distance/similarity estimation.
To allow a fair comparison, we integrated it within \tool as reconstructor. However, unlike autoencoders, \deeproad is computationally very expensive and memory demanding (due to the size of the matrix supplied to PCA). In the paper, authors reduced their training set to 600 images, which were resized to 120x90. During input validation, the three dimensional representation of an online input image is compared to the nominal images by measuring the average of the top-100 minimum distances from the training set.

Our own implementation relaxed the restrictions above by considering a training set consisting of 3,000 randomly sampled images (i.e., $5\times$ improvement w.r.t. the original implementation described in the paper), resized to 224x224, which is the default input size for VGG19~\cite{keras-vgg19}. Keeping the fraction of the training set ($\frac{1}{6}$) constant, we compute similarity based on the average of the top-500 minimal distances.
For implementation details, we made our code publicly available in the replication package accompanying this paper~\cite{tool}.


\subsection{Evaluation of Simulation Results}

To allow a fair comparison with our baseline \deeproad, we evaluated all approaches offline, by splitting the evaluation set of recorded images in \emph{windows of consecutive frames}, which we labelled as either anomalous or normal (\autoref{fig:labelling}).
In anomalous windows, \tool is expected to predict the shortly-following misbehavior. 

\subsubsection{Labelling of Anomaly and Normal Windows in Evaluation Data}\label{sec:labelling}

Let $\boldsymbol{X} = \{x_1, x_2, \ldots, x_n \}$ be the sequence of considered images (frames).
Misbehaviors are represented as $m_j \in \{0,1\}$, where $m_j = 1$ iff a misbehavior is recorded at $x_j \in \boldsymbol{X}$.
We define a healing period as a sequence of $h$ misbehaviour-free frames following a misbehaviour at time $t$.
We define a reaction period as a sequence of $r$ misbehaviour-free frames preceding a misbehaviour at time $t'$ and not intersecting any healing period.
We define an \textit{anomalous window} as $a$ consecutive misbehaviour-free frames followed by a reaction period that does not intersect any healing period.
We define a \textit{normal window} as $b$ consecutive misbehaviour-free frames followed by an anomalous window, or a normal window that does not intersect any healing period.
This is illustrated graphically in \autoref{fig:labelling}. Formally, assuming any misbehavior recorded at time $t \in [1{:}n]$, i.e., $m_t = 1$:

\begin{itemize}
    \item window $x_{t+1}$ to $x_{t+h}$ is labelled as healing period if $m_t = 1$ and $m_j = 0 \: \forall j \in [t+1 {:} t+h]$;
    \item furthermore, the window $x_{t+1}$ to $x_{k-1}$ is also labelled as healing period if $m_k = 1$ with $k > t$ and $k-t-1 < h$, and $m_j = 0 \: \forall j \in [t+1 {:} k-1]$;
    \item window $x_{t-r}$ to $x_{t-1}$ is labelled as reaction period iff $m_t = 1$, $m_j = 0 \: \forall j \in [t-r {:} t-1]$ and no healing period contains any frame from $x_{t-r}$ to $x_{t-1}$;
    \item window $x_{i}$ to $x_{i+a-1}$ is labelled as anomaly window iff a reaction period starts at $x_{i+a}$, $m_j = 0 \: \forall j \in [i {:} i+a-1]$ and no healing period contains any frame from $x_{i}$ to $x_{i+a-1}$;
    \item window $x_{i}$ to $x_{i+b-1}$ is labelled as normal window iff an anomaly or a normal window starts at $x_{i+b}$, $m_j = 0 \: \forall j \in [i {:} i+b-1]$ and no healing period contains any frame from $x_{i}$ to $x_{i+b-1}$.
\end{itemize}

Moreover, if $m_j = 0$ for all $j \in [k{:}n]$, all consecutive windows of size $b$ starting within $x_{k}$ to $x_{n-r-a-1}$ which do not intersect any healing period are labelled as normal.
The labelling described above ensures that after the last misbehavior in a sequence, $h$ \emph{healing images} are ignored (i.e., not labeled) before another anomaly or normal window is defined.
$h$ must be chosen high enough that the car is back safely on the road when the next windows are labelled.
Furthermore, $r$ images occur in between an anomaly window in which the system is supposed to predict the upcoming misbehavior, and the actual misbehavior.
This period would, in practice, be used by the self-healing system to execute countermeasures against the predicted future misbehavior.
Intuitively, misbehavior prediction is expected to be much harder as the value of $r$ increases. In our experiments, we set the value of $n = b = 30$ frames (i.e., normal/anomalous windows), which is $\approx$3s.\footnote{In our setting, Udacity frame rate was approximately 10/12 fps.} The size of the healing window was set to $h = 60$ (> 5s) frames, and the size of the reaction window to $r = 50$ (> 4s) frames.

\subsubsection{Metrics used for Analysis}\label{sec:metrics_analysis}

If the loss score for an image is higher than the automatically estimated threshold $\theta$ (\autoref{sec:usage-scenario}), \tool triggers an alarm.
Consequently, a true positive is defined when \tool triggers an alarm during an anomalous window, early enough to predict a misbehavior. Conversely, a false negative occurs when \tool does not trigger an alarm during an anomalous window, thus failing at predicting a mis-behaviour in time for triggering self-healing.
A false positive represents a false alarm by \tool, whereas true negative cases occur when \tool detects correct detection of normality.


%
%
%
%
%

We assume that a single alarm immediately starts the self healing system, such that multiple consecutive alarms within the healing time have no  effect.
Correspondingly, once a FP occurs and the self healing system is running, additional consecutive FP windows have no effect in practice and are thus excluded from our analysis.

Our goal is to achieve (1)~high recall, or true positive rate (TPR, defined as TP/TP+FN), i.e., true alarms, while (2)~minimizing the complement of specificity, or false positive rate (FPR, defined as FP/TN+FP), i.e., labelling safe situations as unsafe.
We are also interested in F1-score ($F_{1}=2 \cdot \frac{\text{Precision} \times \text{Recall}}{\text{Precision} + \text{Recall}}$) because, in practice, it is informative to have a high F1-score at a given threshold.

We also consider two widely adopted threshold-independent metrics for evaluating classifiers at various thresholds settings such as AUC-ROC (area under the curve of the Receiver Operating Characteristics), and AUC-PRC (area under the Precision-Recall curve).

\subsection{Results}


\begin{table*}[t]
\footnotesize
\setlength{\tabcolsep}{3.3pt}
\renewcommand{\arraystretch}{1}
\centering
\caption{Evaluation Results for all variants of \tool across all SDCs. Best strategies are highlighted.}
\begin{tabular}{@{}lrrrrrrrrrrr@{\hskip 1.5em}rrrrrrrrr@{}}

\toprule

& \multicolumn{2}{c}{}
& \multicolumn{9}{c}{$\epsilon=0.05$, $1-\epsilon=0.95$}
& \multicolumn{9}{c}{$\epsilon=0.01$, $1-\epsilon=0.99$} \\

\cmidrule(r){4-12}
\cmidrule(r){13-21}

& \multicolumn{2}{c}{}
& \multicolumn{8}{c}{\it Unexpected}
& {Nominal}
& \multicolumn{8}{c}{\it Unexpected}
& {Nominal} \\

\cmidrule(r){4-11}
\cmidrule(r){12-12}
\cmidrule(r){13-20}
\cmidrule(r){21-21}


& \text{AUC-PRC$\boldsymbol{\uparrow}$}
& \text{AUC-ROC$\boldsymbol{\uparrow}$}
& \text{TP}
& \text{FP}
& \text{TN}
& \text{FN}
& \text{TPR$\boldsymbol{\uparrow}$}
& \text{FPR$\boldsymbol{\downarrow}$}
& \text{F1}
& \text{Prec.}
& \text{FPR$\boldsymbol{\downarrow}$}
& \text{TP}
& \text{FP}
& \text{TN}
& \text{FN}
& \text{TPR$\boldsymbol{\uparrow}$}
& \text{FPR$\boldsymbol{\downarrow}$}
& \text{F1}
& \text{Prec.}
& \text{FPR$\boldsymbol{\downarrow}$} \\

\midrule
\textit{\textbf{DAVE-2}}&       &       &     &     &        &     &       &       &        &        &       &     &     &        &     &       &       &       &       &       \\
\quad VAE                    & 0.354 & 0.902 & 149 & 304  & 1,970  & 47  & 0.760 & 0.134 & 0.459 & 0.329 & 0.042 & 107 & 183 & 2,959  & 89  & 0.546 & 0.058 & 0.440  & 0.369  & 0.003 \\
\quad DAE                    & 0.330 & 0.891 & 104 & 210  & 2,679  & 92  & 0.531 & 0.073 & 0.408 & 0.331 & 0.041 & 21  & 55  & 4,063  & 175 & 0.107 & 0.013 & 0.154  & 0.276  & 0.010 \\
\quad SAE                    & 0.336 & 0.891 & 138 & 260  & 1,877  & 58  & 0.704 & 0.122 & 0.465 & 0.347 & 0.050 & 108 & 183 & 2,665  & 88  & 0.551 & 0.064 & 0.444  & 0.371  & 0.002 \\
\quad CAE                    & 0.290 & 0.821 & 6   & 22   & 4,208  & 190 & 0.031 & 0.005 & 0.054 & 0.214 & 0.024 & 0   & 0   & 4,282  & 196 & 0 & 0 & n.a. & n.a. & 0 \\
\quad LSTM                   & 0.357 & 0.903 & 16  & 34   & 3,990  & 177 & 0.083 & 0.008 & 0.132 & 0.320 & 0 & 7   & 12  & 4,119  & 186 & 0.036 & 0.003 & 0.066  & 0.368  & 0 \\
\quad \deeproad              & 0.198 & 0.780 & 65  & 344  & 3,170  & 131 & 0.332 & 0.098 & 0.215 & 0.159 & 0.054 & 44  & 250 & 3,651  & 152 & 0.225 & 0.064 & 0.180  & 0.150  & 0.037 \\
\textit{\textbf{Epoch}}      &       &       &     &      &        &     &       &       &       &       &       &     &     &        &     &       &       &        &        &       \\
\quad VAE                    & 0.391 & 0.904 & 158 & 331  & 1,952  & 51  & 0.756 & 0.145 & 0.453 & 0.323 & 0.049 & 106 & 169 & 2,858  & 103 & 0.507 & 0.056 & 0.438  & 0.386  & 0.001 \\
\quad DAE                    & 0.399 & 0.895 & 112 & 188  & 2,720  & 97  & 0.536 & 0.065 & 0.440 & 0.373 & 0.042 & 24  & 34  & 3,653  & 185 & 0.115 & 0.009 & 0.180  & 0.414  & 0.010 \\
\quad SAE                    & 0.386 & 0.883 & 147 & 284  & 2,026  & 62  & 0.703 & 0.123 & 0.459 & 0.341 & 0.050 & 120 & 175 & 2,838  & 89  & 0.574 & 0.058 & 0.476  & 0.407  & 0.002 \\
\quad CAE                    & 0.310 & 0.822 & 6   & 23   & 3,661  & 203 & 0.029 & 0.006 & 0.050 & 0.207 & 0.020 & 0   & 0   & 3,731  & 209 & 0 & 0 & n.a. & n.a. & 0 \\
\quad LSTM                   & 0.385 & 0.879 & 23  & 34   & 3,503  & 175 & 0.116 & 0.010 & 0.180 & 0.404 & 0.001 & 7   & 13  & 3,592  & 191 & 0.035 & 0.004 & 0.064  & 0.350  & 0 \\
\quad \deeproad              & 0.213 & 0.807 & 70  & 308  & 2,917  & 139 & 0.335 & 0.096 & 0.239 & 0.185 & 0.053 & 43  & 201 & 3,240  & 166 & 0.206 & 0.058 & 0.190  & 0.176  & 0.040 \\
\textit{\textbf{Chauffeur}}  &       &       &     &      &        &     &       &       &       &       &       &     &     &        &     &       &       &        &        &       \\
\quad VAE                    & 0.242 & 0.951 & 98  & 392  & 3,700  & 23  & 0.810 & 0.096 & 0.321 & 0.200 & 0.049 & 81  & 267 & 5,391  & 40  & 0.669 & 0.047 & 0.345  & 0.233  & 0.002 \\
\quad DAE                    & 0.203 & 0.944 & 78  & 281  & 5,045  & 43  & 0.645 & 0.053 & 0.325 & 0.217 & 0.051 & 13  & 95  & 7,730  & 108 & 0.107 & 0.012 & 0.114  & 0.120  & 0.009 \\
\quad SAE                    & 0.241 & 0.931 & 96  & 354  & 3,650  & 25  & 0.793 & 0.088 & 0.336 & 0.213 & 0.056 & 86  & 240 & 5,177  & 35  & 0.711 & 0.044 & 0.385  & 0.264  & 0.003 \\
\quad CAE                    & 0.172 & 0.909 & 7   & 34   & 8,127  & 114 & 0.058 & 0.004 & 0.086 & 0.171 & 0.023 & 0   & 0   & 8,229  & 121 & 0 & 0 & n.a. & n.a. & 0 \\
\quad LSTM                   & 0.240 & 0.945 & 11  & 41   & 7,879  & 111 & 0.090 & 0.005 & 0.126 & 0.212 & 0 & 4   & 12  & 8,035  & 118 & 0.033 & 0.002 & 0.058  & 0.250  & 0 \\
\quad \deeproad              & 0.098 & 0.797 & 37  & 594  & 5,564  & 84  & 0.306 & 0.097 & 0.098 & 0.059 & 0.055 & 23  & 458 & 6,551  & 98  & 0.190 & 0.065 & 0.076  & 0.048  & 0.042 \\
\textit{\textbf{Totals}}     &       &       &     &      &        &     &       &       &       &       &       &     &     &        &     &       &       &        &        &       \\
\quad VAE                    & 0.320 & 0.924 & 405 & 1027 & 7,622  & 121 & 0.770 & 0.119 & 0.414 & 0.283 & 0.046 & 294 & 619 & 11,208 & 232 & 0.559 & 0.052 & 0.409  & 0.322  & 0.002 \\
\quad DAE                    & 0.301 & 0.911 & 294 & 679  & 10,444 & 232 & 0.559 & 0.061 & 0.392 & 0.302 & 0.045 & 58  & 184 & 15,446 & 468 & 0.110 & 0.012 & 0.151  & 0.240  & 0.009 \\
\quad SAE                    & 0.312 & 0.907 & 381 & 898  & 7,553  & 145 & 0.724 & 0.106 & 0.422 & 0.298 & 0.052 & 314 & 598 & 10,680 & 212 & 0.597 & 0.053 & 0.437  & 0.344  & 0.002 \\
\quad CAE                    & 0.255 & 0.864 & 19  & 79   & 15,996 & 507 & 0.036 & 0.005 & 0.061 & 0.194 & 0.022 & 0   & 0   & 16,242 & 526 & 0 & 0 & n.a. & n.a. & 0 \\
\quad LSTM                   & 0.329 & 0.915 & 50  & 109  & 15,372 & 463 & 0.098 & 0.007 & 0.149 & 0.315 & 0 & 18  & 37  & 15,746 & 495 & 0.035 & 0.002 & 0.063  & 0.327  & 0 \\
\quad \deeproad              & 0.159 & 0.799 & 172 & 1246 & 11,651 & 354 & 0.327 & 0.097 & 0.177 & 0.121 & 0.054 & 110 & 909 & 13,442 & 416 & 0.209 & 0.063 & 0.142  & 0.108  & 0.040 \\

\bottomrule

\end{tabular}
\label{tab:big_results_table}
\end{table*}

\head{Effectiveness (RQ\textsubscript{1})}
\autoref{tab:big_results_table} presents the effectiveness results on a per-SDC model basis. Columns~2 and~3 show threshold-independent measures of AUC-PRC and AUC-ROC. The remaining of the table shows the effectiveness metrics across two confidence thresholds that we found interesting for analysis and correspond to $\epsilon = 0.05$ and $\epsilon = 0.01$ (at lower values of $\epsilon$, both FPR and TPR get close to zero, making the misbehaviour predictor useless).

Overall, LSTM and VAE are the best performing reconstructors on the AUC-PRC and AUC-ROC metrics.
Columns~12 and~21 (Nominal/FPR) show  FPR under conditions similar to those of the training set. Values are almost always near to zero and occasionally even equal to zero (this is indicated by omitting the decimals) across the variants of \tool. This is an empirical validation of the accuracy of the gamma distribution as a statistical model for the reconstruction errors. In fact, at $\epsilon = 0.05$ most FPR reported in column 12 are very close to the theoretical value, 0.05 (see, e.g., rows under \textit{Totals}). At $\epsilon = 0.01$ some values drop to zero. This means that in those configurations \tool will raise no false alarm when the SDC drives in nominal conditions.
For instance, with both thresholds, LSTM never raised false alarms within the 23,728 considered normal windows .


In terms of TPR (to be maximized) and FPR (to be minimized), the best reconstructors are VAE and SAE, with comparable overall performance: 
 77\%, 11\% (TPR, FPR of VAE) and 72\%, 10\% (SAE) at $\epsilon = 0.05$; 55\%, 5\% (VAE)  and 59\%, 5\% (SAE) at $\epsilon = 0.01$.
It can be noticed that FPR is higher than $\epsilon$ (10\% vs 5\% and 5\% vs 1\%) with both reconstructors. This is expected, since we are measuring FPR in tracks with injected anomalies. These tracks differ substantially from the nominal tracks even in conditions not so extreme as to cause a misbehaviour, hence inflating the FPR a bit. However, we can notice that even in such non nominal conditions, the FPR remains low and not too far from the theoretical prediction $\epsilon$.



We can \textbf{answer to RQ1} by noticing that with reconstructors VAE and SAE we achieve a FPR in nominal conditions close to the theoretical expectations (resp. 5\% and 1\%, in the two considered configurations); that in anomalous conditions FPR increases by a moderate amount (resp., +5\% and +4\%); and that the achieved TPR is quite high (with VAE, SAE, resp. 77\%, 72\% and 55\%, 59\%).

\begin{figure}[t]
    \includegraphics[scale=0.25]{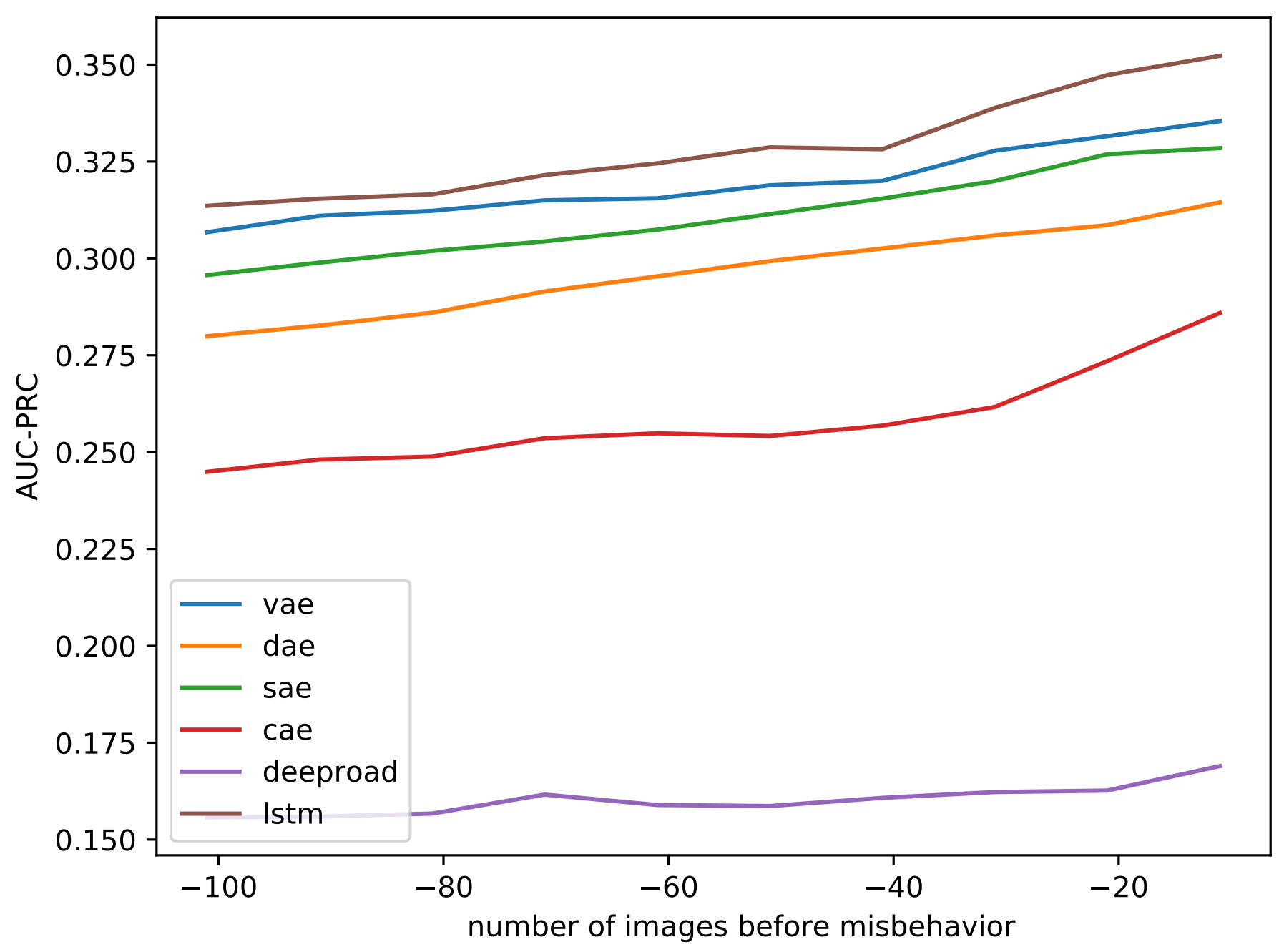}
    \caption{Misbehavior prediction capability over time.}
    \label{fig:prc_reaction}
\end{figure}

\head{Prediction (RQ\textsubscript{2})}
\autoref{fig:prc_reaction} shows the PRC-AUC of the various configurations of \tool over different reaction periods.
The general trend is that predictions get harder when the SDC is far from a critical scenario, having a longer reaction period to prevent the misbehavior, but quite surprisingly there is no drop in performance as we move away from the misbehaviour. Our explanation of this unexpected finding is that the tracks used for the evaluation of the approach contain always a relatively high degree of anomalous features, which might trigger a self-healing reaction. Occasionally, the level of detected anomalies  surpasses the threshold and the misbehaviour predictor raises an alarm. Correspondingly, although slightly reduced,  the signals of an upcoming misbehaviour exist in images quite far (even 60 frames or around 6s) from the misbehaviour. 

We can \textbf{answer to RQ2} by noticing that the performance of \tool degrades smoothly as we anticipate the prediction (AUC-PRC remains quite high even 6s before the misbehaviour), but we should also remark that this result must be taken with care and might be partially due to the characteristics of the considered tracks, which contain a continuously and smoothly increasing degree of injected anomalies by design.



\head{Comparison (RQ\textsubscript{3})}
%
%
In our experiments, \tool is constantly superior to \deeproad at predicting misbehaviours. Results of AUC-PRC and AUC-ROC show significant improvements across all thresholds, and regardless of the technique being used and the reaction period considered (in \autoref{fig:prc_reaction}, \deeproad is the lowest curve).
With $\epsilon = 0.05$ (resp. $\epsilon = 0.01$), 
VAE and SAE (see \autoref{tab:big_results_table}) expose more than twice the misbehaviours exposed by \deeproad, with a TPR =  77\%, 72\% vs 32\% (resp., 55\%, 59\% vs 20\%), at comparable false positive rate FPR = 11\%, 10\% vs 9\% (resp., 5\% vs 6\%).

While reimplementing \deeproad, we noticed  its high computational cost that makes it quite unsuitable for online misbehaviour prediction.
SDC manufacturers use much larger training datasets than the one we used for our empirical study and differently from autoencoders such as VAE and SAE, \deeproad is quite sensitive to the size of the training set, which must be sub-sampled dramatically to make the approach applicable. On the contrary, autoencoders seem a very promising option, given their relatively simple architecture. In particular, SAE is very efficient, yet it has comparable performance as the more sophisticated VAE.
Hence, as an input validation framework, \deeproad has shown to be both computationally very expensive and inaccurate, which makes it less promising and desirable than \tool for online misbehaviour prediction.

Our \textbf{answer to RQ3} is that \tool outperforms \deeproad in all respects: computational cost, accuracy of misbehaviour prediction (see TPR) and minimization of false alarms (see FPR and AUC-PRC).

\subsection{Threats to Validity}\label{sec:ttv}

\head{Internal validity}
We compared all variants of \tool and \deeproad under identical parameter settings, and on the same evaluation set.
The main threat to internal validity concerns our custom implementation of unexpected conditions within the simulator. However, this was a mandatory choice, since we are not aware of open source driving simulators that can inject unexpected execution contexts in a controllable way. 
Another possible threat may be the choice and the training of our own SDCs, which may exhibit a large number of misbehaviour if trained inadequately. 
We mitigated this threat by training and fine-tuning the best publicly available driving models. Our own implementation of \deeproad may be another threat to internal validity, that we mitigated by developing an implementation which improves the original one by processing $5\times$ more information.

\head{External validity}
We used a limited number of self-driving systems in our evaluation, as well as tracks, which pose a threat in terms of generalizability of our results. We tried to mitigate this threat by choosing popular subjects developed with real-world frameworks (e..g, Keras). 

\head{Reproducibility} 
All our results, the source code of \tool, the simulator, and all subjects are available~\cite{tool}, making the evaluation repeatable and our results reproducible.

\section{Related Work}\label{sec:relwork}


\head{Adversarial Input Generation}
Adversarial generation approaches aim at generating inputs that trigger inconsistencies between multiple autonomous driving systems~\cite{deepxplore}, or between the original and transformed driving scenarios~\cite{deeptest,deeproad,7313519}. 
%
%
These works exploit the well-known fragility of DNNs to adversarial examples. Therefore, their main use-case concerns the identification of underrepresented scenarios in the training data (e.g., snowy weather condition) to support re-training and better generalization after re-training. 
The only comparable technique is the online input validation of \deeproad~\cite{deeproad}, for which we carried out an explicit comparison in our empirical study, finding poor performance when used for online misbehaviour prediction.

Despite the different goal (test generation vs misbehaviour prediction), we share with these works the problem of how to empirically validate the proposed technique in the absence of a precise oracle that defines the expected behaviour of a self-driving car. The prevalent choice in test generators~\cite{deeptest,deeproad,7313519,deepxplore} is to address the oracle problem by \textit{differential testing}, i.e., by comparing the behaviours of multiple DNNs, or by \textit{metamorphic testing}, i.e., by comparing the behaviour before and after applying a metamorphic transformation to the input. Approaches based on verification are also being under investigation~\cite{deepsafe}.
In this paper, we adopt a precise definition of \textit{DNN misbehaviour}, which gives us a very accurate \textit{functional oracle}, with no need for differential testing, metamorphic testing, or verification. 


\head{Search-based Generation}
Abdessalem et al.~\cite{Abdessalem-ASE18-1,Abdessalem-ASE18-2,Abdessalem-ICSE18} combine genetic algorithms and machine learning  to test a pedestrian detection system. 
Mullins et al.~\cite{MULLINS2018197} use Gaussian processes to drive the search towards yet unexplored regions of the input space, whereas Gambi et al.~\cite{Gambi:2019:ATS:3293882.3330566} propose \textsc{AsFault}, a search-based test generator for autonomous vehicles based on procedural content generation. 
\textsc{AsFault} uses search operators which mutate and recombine road segments to construct road networks for testing the lane keeping functionality of self-driving cars. Their goal is to generate extreme and challenging roads, maximizing the number of observed OBEs, while our goal is to avoid OBEs by predicting misbehaviours.


\head{Anomaly Detection for Time Series}
Anomaly detection for self-driving vehicles has been studied from a security perspective. For example, Narayanan et al.~\cite{Narayanan} use a Hidden Markov Model to detect malicious behaviours from real vehicles, and issue alerts while a vehicle is in operation. Taylor et al.~\cite{7796898} use LSTM neural networks to detect car's controller area network (CAN) bus attacks, whereas Marchetti et al.~\cite{7740627} used an information theoretic approach.
Lin et al.~\cite{5509781} used the Mahalanobis distance between multiple sensor data to identify unusual events in Unmanned Aerial Vehicles (UAVs). However, their approach is not image-based and thus is not comparable with our work.
Moreover, ours is the first approach that detects unexpected driving conditions for the prediction of misbehaviours of an autonomous vehicle to enable self-healing.

\section{Conclusions and Future Work}\label{sec:conclusions}

In this paper, we studied the problem of estimating the confidence of the DNN-based autonomous in response to unexpected execution contexts.  
Our tool \tool was able to anticipate by several seconds many potentially safety-critical misbehaviours, such as out of bound episodes or collisions, with a low false alarm rate, outperforming the input validator of \deeproad. 
Future work concerns devising novel metrics of DNN confidence, including white-box ones, with a potential for hybridization. 
It would be also interesting to characterize and predict other kind of misbehaviours (e.g., derivative of steering angle) as well as implementing confidence-guided self-healing within the simulator.
We believe that our promising results in online misbehaviour detection, united with the availability of a labeled dataset of crashes and a simulation environment, can foster novel approaches for online prediction and self-healing of autonomous driving systems.



\balance
\bibliographystyle{ACM-Reference-Format}
\bibliography{paper}


\begin{thebibliography}{42}


\ifx \showCODEN    \undefined \def \showCODEN     #1{\unskip}     \fi
\ifx \showDOI      \undefined \def \showDOI       #1{#1}\fi
\ifx \showISBNx    \undefined \def \showISBNx     #1{\unskip}     \fi
\ifx \showISBNxiii \undefined \def \showISBNxiii  #1{\unskip}     \fi
\ifx \showISSN     \undefined \def \showISSN      #1{\unskip}     \fi
\ifx \showLCCN     \undefined \def \showLCCN      #1{\unskip}     \fi
\ifx \shownote     \undefined \def \shownote      #1{#1}          \fi
\ifx \showarticletitle \undefined \def \showarticletitle #1{#1}   \fi
\ifx \showURL      \undefined \def \showURL       {\relax}        \fi
\providecommand\bibfield[2]{#2}
\providecommand\bibinfo[2]{#2}
\providecommand\natexlab[1]{#1}
\providecommand\showeprint[2][]{arXiv:#2}

\bibitem[\protect\citeauthoryear{Abdessalem, Panichella, Nejati, Briand, and
  Stifter}{Abdessalem et~al\mbox{.}}{2018}]%
        {Abdessalem-ASE18-1}
\bibfield{author}{\bibinfo{person}{Raja~Ben Abdessalem},
  \bibinfo{person}{Annibale Panichella}, \bibinfo{person}{Shiva Nejati},
  \bibinfo{person}{Lionel~C. Briand}, {and} \bibinfo{person}{Thomas Stifter}.}
  \bibinfo{year}{2018}\natexlab{}.
\newblock \showarticletitle{Testing Autonomous Cars for Feature Interaction
  Failures Using Many-objective Search}. In
  \bibinfo{booktitle}{\emph{Proceedings of the 33rd ACM/IEEE International
  Conference on Automated Software Engineering}} \emph{(\bibinfo{series}{ASE
  2018})}. \bibinfo{publisher}{ACM}, \bibinfo{address}{New York, NY, USA},
  \bibinfo{pages}{143--154}.
\newblock
\showISBNx{978-1-4503-5937-5}
\urldef\tempurl%
\url{https://doi.org/10.1145/3238147.3238192}
\showDOI{\tempurl}


\bibitem[\protect\citeauthoryear{Alvarez{-}Melis and Jaakkola}{Alvarez{-}Melis
  and Jaakkola}{2018}]%
        {AlvarezMelisJ18}
\bibfield{author}{\bibinfo{person}{David Alvarez{-}Melis} {and}
  \bibinfo{person}{Tommi~S. Jaakkola}.} \bibinfo{year}{2018}\natexlab{}.
\newblock \showarticletitle{Towards Robust Interpretability with
  Self-Explaining Neural Networks}. In \bibinfo{booktitle}{\emph{Annual
  Conference on Neural Information Processing Systems (NeurIPS)}}.
  \bibinfo{pages}{7786--7795}.
\newblock


\bibitem[\protect\citeauthoryear{An and Cho}{An and Cho}{2015}]%
        {An2015VariationalAB}
\bibfield{author}{\bibinfo{person}{Jinwon An} {and} \bibinfo{person}{Sungzoon
  Cho}.} \bibinfo{year}{2015}\natexlab{}.
\newblock \showarticletitle{Variational Autoencoder based Anomaly Detection
  using Reconstruction Probability}.
\newblock


\bibitem[\protect\citeauthoryear{{Ben Abdessalem}, {Nejati}, {Briand}, and
  {Stifter}}{{Ben Abdessalem} et~al\mbox{.}}{2016}]%
        {Abdessalem-ASE18-2}
\bibfield{author}{\bibinfo{person}{R. {Ben Abdessalem}}, \bibinfo{person}{S.
  {Nejati}}, \bibinfo{person}{L.~C. {Briand}}, {and} \bibinfo{person}{T.
  {Stifter}}.} \bibinfo{year}{2016}\natexlab{}.
\newblock \showarticletitle{Testing advanced driver assistance systems using
  multi-objective search and neural networks}. In
  \bibinfo{booktitle}{\emph{2016 31st IEEE/ACM International Conference on
  Automated Software Engineering (ASE)}}. \bibinfo{pages}{63--74}.
\newblock


\bibitem[\protect\citeauthoryear{{Ben Abdessalem}, {Nejati}, {C. Briand}, and
  {Stifter}}{{Ben Abdessalem} et~al\mbox{.}}{2018}]%
        {Abdessalem-ICSE18}
\bibfield{author}{\bibinfo{person}{R. {Ben Abdessalem}}, \bibinfo{person}{S.
  {Nejati}}, \bibinfo{person}{L. {C. Briand}}, {and} \bibinfo{person}{T.
  {Stifter}}.} \bibinfo{year}{2018}\natexlab{}.
\newblock \showarticletitle{Testing Vision-Based Control Systems Using
  Learnable Evolutionary Algorithms}. In \bibinfo{booktitle}{\emph{2018
  IEEE/ACM 40th International Conference on Software Engineering (ICSE)}}.
  \bibinfo{pages}{1016--1026}.
\newblock
\urldef\tempurl%
\url{https://doi.org/10.1145/3180155.3180160}
\showDOI{\tempurl}


\bibitem[\protect\citeauthoryear{{BGR Media, LLC}}{{BGR Media, LLC}}{2018}]%
        {10-million-miles}
\bibfield{author}{\bibinfo{person}{{BGR Media, LLC}}.}
  \bibinfo{year}{2018}\natexlab{}.
\newblock \bibinfo{title}{{Waymo's self-driving cars hit 10 million miles}}.
\newblock
  \bibinfo{howpublished}{\url{https://techcrunch.com/2018/10/10/waymos-self-driving-cars-hit-10-million-miles}}.
\newblock
\newblock
\shownote{Online; accessed 18 August 2019.}


\bibitem[\protect\citeauthoryear{Bojarski, Testa, Dworakowski, Firner, Flepp,
  Goyal, Jackel, Monfort, Muller, Zhang, Zhang, Zhao, and Zieba}{Bojarski
  et~al\mbox{.}}{2016}]%
        {nvidia-dave2}
\bibfield{author}{\bibinfo{person}{Mariusz Bojarski},
  \bibinfo{person}{Davide~Del Testa}, \bibinfo{person}{Daniel Dworakowski},
  \bibinfo{person}{Bernhard Firner}, \bibinfo{person}{Beat Flepp},
  \bibinfo{person}{Prasoon Goyal}, \bibinfo{person}{Lawrence~D. Jackel},
  \bibinfo{person}{Mathew Monfort}, \bibinfo{person}{Urs Muller},
  \bibinfo{person}{Jiakai Zhang}, \bibinfo{person}{Xin Zhang},
  \bibinfo{person}{Jake Zhao}, {and} \bibinfo{person}{Karol Zieba}.}
  \bibinfo{year}{2016}\natexlab{}.
\newblock \showarticletitle{End to End Learning for Self-Driving Cars.}
\newblock \bibinfo{journal}{\emph{CoRR}}  \bibinfo{volume}{abs/1604.07316}
  (\bibinfo{year}{2016}).
\newblock
\urldef\tempurl%
\url{http://arxiv.org/abs/1604.07316}
\showURL{%
\tempurl}


\bibitem[\protect\citeauthoryear{Burkhard, Vos, Munzinger, Enders, and
  Schramm}{Burkhard et~al\mbox{.}}{2018}]%
        {10.1007/978-3-658-21194-3_53}
\bibfield{author}{\bibinfo{person}{Georg Burkhard}, \bibinfo{person}{S. Vos},
  \bibinfo{person}{N. Munzinger}, \bibinfo{person}{E. Enders}, {and}
  \bibinfo{person}{D. Schramm}.} \bibinfo{year}{2018}\natexlab{}.
\newblock \showarticletitle{Requirements on driving dynamics in autonomous
  driving with regard to motion and comfort}. In \bibinfo{booktitle}{\emph{18.
  Internationales Stuttgarter Symposium}},
  \bibfield{editor}{\bibinfo{person}{Michael Bargende},
  \bibinfo{person}{Hans-Christian Reuss}, {and} \bibinfo{person}{Jochen
  Wiedemann}} (Eds.). \bibinfo{publisher}{Springer Fachmedien Wiesbaden},
  \bibinfo{address}{Wiesbaden}, \bibinfo{pages}{683--697}.
\newblock


\bibitem[\protect\citeauthoryear{Campos, Zimek, Sander, Campello,
  Micenkov\'{a}, Schubert, Assent, and Houle}{Campos et~al\mbox{.}}{2016}]%
        {Campos:2016:EUO:2962863.2962870}
\bibfield{author}{\bibinfo{person}{Guilherme~O. Campos},
  \bibinfo{person}{Arthur Zimek}, \bibinfo{person}{J\"{o}rg Sander},
  \bibinfo{person}{Ricardo~J. Campello}, \bibinfo{person}{Barbora
  Micenkov\'{a}}, \bibinfo{person}{Erich Schubert}, \bibinfo{person}{Ira
  Assent}, {and} \bibinfo{person}{Michael~E. Houle}.}
  \bibinfo{year}{2016}\natexlab{}.
\newblock \showarticletitle{On the Evaluation of Unsupervised Outlier
  Detection: Measures, Datasets, and an Empirical Study}.
\newblock \bibinfo{journal}{\emph{Data Min. Knowl. Discov.}}
  \bibinfo{volume}{30}, \bibinfo{number}{4} (\bibinfo{date}{July}
  \bibinfo{year}{2016}), \bibinfo{pages}{891--927}.
\newblock
\showISSN{1384-5810}
\urldef\tempurl%
\url{https://doi.org/10.1007/s10618-015-0444-8}
\showDOI{\tempurl}


\bibitem[\protect\citeauthoryear{Cerf}{Cerf}{2018}]%
        {Cerf:2018:CSC:3181977.3177753}
\bibfield{author}{\bibinfo{person}{Vinton~G. Cerf}.}
  \bibinfo{year}{2018}\natexlab{}.
\newblock \showarticletitle{A Comprehensive Self-driving Car Test}.
\newblock \bibinfo{journal}{\emph{Commun. ACM}} \bibinfo{volume}{61},
  \bibinfo{number}{2} (\bibinfo{date}{Jan.} \bibinfo{year}{2018}),
  \bibinfo{pages}{7--7}.
\newblock
\showISSN{0001-0782}
\urldef\tempurl%
\url{https://doi.org/10.1145/3177753}
\showDOI{\tempurl}


\bibitem[\protect\citeauthoryear{{Electrek}}{{Electrek}}{2016a}]%
        {tesla-accident-3}
\bibfield{author}{\bibinfo{person}{{Electrek}}.}
  \bibinfo{year}{2016}\natexlab{a}.
\newblock \bibinfo{title}{{A Google self-driving car caused a crash for the
  first time}}.
\newblock
  \bibinfo{howpublished}{\url{https://electrek.co/2016/07/01/understanding-fatal-tesla-accident-autopilot-nhtsa-probe/}}.
\newblock
\newblock
\shownote{Online; accessed 18 August 2019.}


\bibitem[\protect\citeauthoryear{{Electrek}}{{Electrek}}{2016b}]%
        {tesla-accident-4}
\bibfield{author}{\bibinfo{person}{{Electrek}}.}
  \bibinfo{year}{2016}\natexlab{b}.
\newblock \bibinfo{title}{{Tesla Model S driver crashes into a van while on
  Autopilot}}.
\newblock
  \bibinfo{howpublished}{\url{https://electrek.co/2016/05/26/tesla-model-s-crash-autopilot-video/}}.
\newblock
\newblock
\shownote{Online; accessed 18 August 2019.}


\bibitem[\protect\citeauthoryear{Gambi, Mueller, and Fraser}{Gambi
  et~al\mbox{.}}{2019}]%
        {Gambi:2019:ATS:3293882.3330566}
\bibfield{author}{\bibinfo{person}{Alessio Gambi}, \bibinfo{person}{Marc
  Mueller}, {and} \bibinfo{person}{Gordon Fraser}.}
  \bibinfo{year}{2019}\natexlab{}.
\newblock \showarticletitle{Automatically Testing Self-driving Cars with
  Search-based Procedural Content Generation}. In
  \bibinfo{booktitle}{\emph{Proceedings of the 28th ACM SIGSOFT International
  Symposium on Software Testing and Analysis}} \emph{(\bibinfo{series}{ISSTA
  2019})}. \bibinfo{publisher}{ACM}, \bibinfo{address}{New York, NY, USA},
  \bibinfo{pages}{318--328}.
\newblock
\showISBNx{978-1-4503-6224-5}
\urldef\tempurl%
\url{https://doi.org/10.1145/3293882.3330566}
\showDOI{\tempurl}


\bibitem[\protect\citeauthoryear{Gopinath, Katz, P{\u{a}}s{\u{a}}reanu, and
  Barrett}{Gopinath et~al\mbox{.}}{2018}]%
        {deepsafe}
\bibfield{author}{\bibinfo{person}{Divya Gopinath}, \bibinfo{person}{Guy Katz},
  \bibinfo{person}{Corina~S. P{\u{a}}s{\u{a}}reanu}, {and}
  \bibinfo{person}{Clark Barrett}.} \bibinfo{year}{2018}\natexlab{}.
\newblock \showarticletitle{DeepSafe: A Data-Driven Approach for Assessing
  Robustness of Neural Networks}. In \bibinfo{booktitle}{\emph{Automated
  Technology for Verification and Analysis}},
  \bibfield{editor}{\bibinfo{person}{Shuvendu~K. Lahiri} {and}
  \bibinfo{person}{Chao Wang}} (Eds.). \bibinfo{publisher}{Springer
  International Publishing}, \bibinfo{address}{Cham}, \bibinfo{pages}{3--19}.
\newblock
\showISBNx{978-3-030-01090-4}


\bibitem[\protect\citeauthoryear{Hochreiter and Schmidhuber}{Hochreiter and
  Schmidhuber}{1997}]%
        {Hochreiter:1997:LSM:1246443.1246450}
\bibfield{author}{\bibinfo{person}{Sepp Hochreiter} {and}
  \bibinfo{person}{J\"{u}rgen Schmidhuber}.} \bibinfo{year}{1997}\natexlab{}.
\newblock \showarticletitle{Long Short-Term Memory}.
\newblock \bibinfo{journal}{\emph{Neural Comput.}} \bibinfo{volume}{9},
  \bibinfo{number}{8} (\bibinfo{date}{Nov.} \bibinfo{year}{1997}),
  \bibinfo{pages}{1735--1780}.
\newblock
\showISSN{0899-7667}
\urldef\tempurl%
\url{https://doi.org/10.1162/neco.1997.9.8.1735}
\showDOI{\tempurl}


\bibitem[\protect\citeauthoryear{{keras}}{{keras}}{[n.d.]}]%
        {keras-autoencoders}
\bibfield{author}{\bibinfo{person}{{keras}}.}
  \bibinfo{year}{[n.d.]}\natexlab{}.
\newblock \bibinfo{title}{{Building Autoencoders in Keras}}.
\newblock
  \bibinfo{howpublished}{\url{https://blog.keras.io/building-autoencoders-in-keras.html}}.
\newblock
\newblock
\shownote{Online; accessed 21 August 2019.}


\bibitem[\protect\citeauthoryear{{Keras}}{{Keras}}{[n.d.]}]%
        {keras-vgg19}
\bibfield{author}{\bibinfo{person}{{Keras}}.}
  \bibinfo{year}{[n.d.]}\natexlab{}.
\newblock \bibinfo{title}{{VGG19}}.
\newblock \bibinfo{howpublished}{\url{https://keras.io/applications/\#vgg19/}}.
\newblock
\newblock
\shownote{Online; accessed 21 August 2019.}


\bibitem[\protect\citeauthoryear{Kim, Feldt, and Yoo}{Kim
  et~al\mbox{.}}{2019}]%
        {Kim:2019:GDL:3339505.3339634}
\bibfield{author}{\bibinfo{person}{Jinhan Kim}, \bibinfo{person}{Robert Feldt},
  {and} \bibinfo{person}{Shin Yoo}.} \bibinfo{year}{2019}\natexlab{}.
\newblock \showarticletitle{Guiding Deep Learning System Testing Using Surprise
  Adequacy}. In \bibinfo{booktitle}{\emph{Proceedings of the 41st International
  Conference on Software Engineering}} \emph{(\bibinfo{series}{ICSE '19})}.
  \bibinfo{publisher}{IEEE Press}, \bibinfo{address}{Piscataway, NJ, USA},
  \bibinfo{pages}{1039--1049}.
\newblock
\urldef\tempurl%
\url{https://doi.org/10.1109/ICSE.2019.00108}
\showDOI{\tempurl}


\bibitem[\protect\citeauthoryear{{Lin}, {Khalastchi}, and {Kaminka}}{{Lin}
  et~al\mbox{.}}{2010}]%
        {5509781}
\bibfield{author}{\bibinfo{person}{R. {Lin}}, \bibinfo{person}{E.
  {Khalastchi}}, {and} \bibinfo{person}{G.~A. {Kaminka}}.}
  \bibinfo{year}{2010}\natexlab{}.
\newblock \showarticletitle{Detecting anomalies in unmanned vehicles using the
  Mahalanobis distance}. In \bibinfo{booktitle}{\emph{2010 IEEE International
  Conference on Robotics and Automation}}. \bibinfo{pages}{3038--3044}.
\newblock
\showISSN{1050-4729}
\urldef\tempurl%
\url{https://doi.org/10.1109/ROBOT.2010.5509781}
\showDOI{\tempurl}


\bibitem[\protect\citeauthoryear{{Marchetti}, {Stabili}, {Guido}, and
  {Colajanni}}{{Marchetti} et~al\mbox{.}}{2016}]%
        {7740627}
\bibfield{author}{\bibinfo{person}{M. {Marchetti}}, \bibinfo{person}{D.
  {Stabili}}, \bibinfo{person}{A. {Guido}}, {and} \bibinfo{person}{M.
  {Colajanni}}.} \bibinfo{year}{2016}\natexlab{}.
\newblock \showarticletitle{Evaluation of anomaly detection for in-vehicle
  networks through information-theoretic algorithms}. In
  \bibinfo{booktitle}{\emph{2016 IEEE 2nd International Forum on Research and
  Technologies for Society and Industry Leveraging a better tomorrow (RTSI)}}.
  \bibinfo{pages}{1--6}.
\newblock
\urldef\tempurl%
\url{https://doi.org/10.1109/RTSI.2016.7740627}
\showDOI{\tempurl}


\bibitem[\protect\citeauthoryear{Masci, Meier, Cire{\c{s}}an, and
  Schmidhuber}{Masci et~al\mbox{.}}{2011}]%
        {10.1007/978-3-642-21735-7_7}
\bibfield{author}{\bibinfo{person}{Jonathan Masci}, \bibinfo{person}{Ueli
  Meier}, \bibinfo{person}{Dan Cire{\c{s}}an}, {and}
  \bibinfo{person}{J{\"u}rgen Schmidhuber}.} \bibinfo{year}{2011}\natexlab{}.
\newblock \showarticletitle{Stacked Convolutional Auto-Encoders for
  Hierarchical Feature Extraction}. In \bibinfo{booktitle}{\emph{Artificial
  Neural Networks and Machine Learning -- ICANN 2011}},
  \bibfield{editor}{\bibinfo{person}{Timo Honkela},
  \bibinfo{person}{W{\l}odzis{\l}aw Duch}, \bibinfo{person}{Mark Girolami},
  {and} \bibinfo{person}{Samuel Kaski}} (Eds.). \bibinfo{publisher}{Springer
  Berlin Heidelberg}, \bibinfo{address}{Berlin, Heidelberg},
  \bibinfo{pages}{52--59}.
\newblock


\bibitem[\protect\citeauthoryear{Mullins, Stankiewicz, Hawthorne, and
  Gupta}{Mullins et~al\mbox{.}}{2018}]%
        {MULLINS2018197}
\bibfield{author}{\bibinfo{person}{Galen~E. Mullins}, \bibinfo{person}{Paul~G.
  Stankiewicz}, \bibinfo{person}{R.~Chad Hawthorne}, {and}
  \bibinfo{person}{Satyandra~K. Gupta}.} \bibinfo{year}{2018}\natexlab{}.
\newblock \showarticletitle{Adaptive generation of challenging scenarios for
  testing and evaluation of autonomous vehicles}.
\newblock \bibinfo{journal}{\emph{Journal of Systems and Software}}
  \bibinfo{volume}{137} (\bibinfo{year}{2018}), \bibinfo{pages}{197 -- 215}.
\newblock
\showISSN{0164-1212}
\urldef\tempurl%
\url{https://doi.org/10.1016/j.jss.2017.10.031}
\showDOI{\tempurl}


\bibitem[\protect\citeauthoryear{{Müller}, {Hospach}, {Bringmann}, {Gerlach},
  and {Rosenstiel}}{{Müller} et~al\mbox{.}}{2015}]%
        {7313519}
\bibfield{author}{\bibinfo{person}{S. {Müller}}, \bibinfo{person}{D.
  {Hospach}}, \bibinfo{person}{O. {Bringmann}}, \bibinfo{person}{J. {Gerlach}},
  {and} \bibinfo{person}{W. {Rosenstiel}}.} \bibinfo{year}{2015}\natexlab{}.
\newblock \showarticletitle{Robustness Evaluation and Improvement for
  Vision-Based Advanced Driver Assistance Systems}. In
  \bibinfo{booktitle}{\emph{2015 IEEE 18th International Conference on
  Intelligent Transportation Systems}}. \bibinfo{pages}{2659--2664}.
\newblock
\showISSN{2153-0017}
\urldef\tempurl%
\url{https://doi.org/10.1109/ITSC.2015.427}
\showDOI{\tempurl}


\bibitem[\protect\citeauthoryear{Narayanan, Mittal, and Joshi}{Narayanan
  et~al\mbox{.}}{2016}]%
        {Narayanan}
\bibfield{author}{\bibinfo{person}{Sandeep~Nair Narayanan},
  \bibinfo{person}{Sudip Mittal}, {and} \bibinfo{person}{Anupam Joshi}.}
  \bibinfo{year}{2016}\natexlab{}.
\newblock \showarticletitle{OBD SecureAlert: An Anomaly Detection System for
  Vehicles}. In \bibinfo{booktitle}{\emph{IEEE Workshop on Smart Service
  Systems (SmartSys 2016)}}.
\newblock


\bibitem[\protect\citeauthoryear{particle-system}{particle-system}{2019}]%
        {particle-system}
particle-system \bibinfo{year}{2019}\natexlab{}.
\newblock \bibinfo{title}{Unity3d Particle System.}
\newblock
  \bibinfo{howpublished}{\url{https://docs.unity3d.com/ScriptReference/ParticleSystem.html}}.
\newblock


\bibitem[\protect\citeauthoryear{Pei, Cao, Yang, and Jana}{Pei
  et~al\mbox{.}}{2017}]%
        {deepxplore}
\bibfield{author}{\bibinfo{person}{Kexin Pei}, \bibinfo{person}{Yinzhi Cao},
  \bibinfo{person}{Junfeng Yang}, {and} \bibinfo{person}{Suman Jana}.}
  \bibinfo{year}{2017}\natexlab{}.
\newblock \showarticletitle{DeepXplore: Automated Whitebox Testing of Deep
  Learning Systems}. In \bibinfo{booktitle}{\emph{Proceedings of the 26th
  Symposium on Operating Systems Principles}} \emph{(\bibinfo{series}{SOSP
  '17})}. \bibinfo{publisher}{ACM}, \bibinfo{address}{New York, NY, USA},
  \bibinfo{pages}{1--18}.
\newblock
\showISBNx{978-1-4503-5085-3}
\urldef\tempurl%
\url{https://doi.org/10.1145/3132747.3132785}
\showDOI{\tempurl}


\bibitem[\protect\citeauthoryear{Shannon}{Shannon}{1948}]%
        {Shannon1948}
\bibfield{author}{\bibinfo{person}{Claude~Elwood Shannon}.}
  \bibinfo{year}{1948}\natexlab{}.
\newblock \showarticletitle{A Mathematical Theory of Communication}.
\newblock \bibinfo{journal}{\emph{The Bell System Technical Journal}}
  \bibinfo{volume}{27}, \bibinfo{number}{3} (\bibinfo{date}{7}
  \bibinfo{year}{1948}), \bibinfo{pages}{379--423}.
\newblock
\urldef\tempurl%
\url{https://doi.org/10.1002/j.1538-7305.1948.tb01338.x}
\showDOI{\tempurl}


\bibitem[\protect\citeauthoryear{Simonyan and Zisserman}{Simonyan and
  Zisserman}{[n.d.]}]%
        {Simonyan2014}
\bibfield{author}{\bibinfo{person}{Karen Simonyan} {and}
  \bibinfo{person}{Andrew Zisserman}.} \bibinfo{year}{[n.d.]}\natexlab{}.
\newblock \showarticletitle{Very Deep Convolutional Networks for Large-Scale
  Image Recognition}.
\newblock  (\bibinfo{year}{[n.\,d.]}).
\newblock
\showeprint[arXiv]{cs.CV/1409.1556v6}
\newblock
\shownote{VGGNet, https://gist.github.com/baraldilorenzo/07d7802847aaad0a35d3.}


\bibitem[\protect\citeauthoryear{{Taylor}, {Leblanc}, and {Japkowicz}}{{Taylor}
  et~al\mbox{.}}{2016}]%
        {7796898}
\bibfield{author}{\bibinfo{person}{A. {Taylor}}, \bibinfo{person}{S.
  {Leblanc}}, {and} \bibinfo{person}{N. {Japkowicz}}.}
  \bibinfo{year}{2016}\natexlab{}.
\newblock \showarticletitle{Anomaly Detection in Automobile Control Network
  Data with Long Short-Term Memory Networks}. In \bibinfo{booktitle}{\emph{2016
  IEEE International Conference on Data Science and Advanced Analytics
  (DSAA)}}. \bibinfo{pages}{130--139}.
\newblock
\urldef\tempurl%
\url{https://doi.org/10.1109/DSAA.2016.20}
\showDOI{\tempurl}


\bibitem[\protect\citeauthoryear{{Team Chauffeur}}{{Team Chauffeur}}{2016}]%
        {chauffeur}
\bibfield{author}{\bibinfo{person}{{Team Chauffeur}}.}
  \bibinfo{year}{2016}\natexlab{}.
\newblock \bibinfo{title}{{Steering angle model: Chauffeur}}.
\newblock
  \bibinfo{howpublished}{\url{https://github.com/udacity/self-driving-car/tree/master/steering-models/community-models/chauffeur}}.
\newblock
\newblock
\shownote{Online; accessed 18 August 2019.}


\bibitem[\protect\citeauthoryear{{Team Epoch}}{{Team Epoch}}{2016}]%
        {epoch}
\bibfield{author}{\bibinfo{person}{{Team Epoch}}.}
  \bibinfo{year}{2016}\natexlab{}.
\newblock \bibinfo{title}{{Steering angle model: Epoch}}.
\newblock
  \bibinfo{howpublished}{\url{https://github.com/udacity/self-driving-car/tree/master/steering-models/community-models/cg23}}.
\newblock
\newblock
\shownote{Online; accessed 18 August 2019.}


\bibitem[\protect\citeauthoryear{{The Verge}}{{The Verge}}{2016}]%
        {tesla-accident-2}
\bibfield{author}{\bibinfo{person}{{The Verge}}.}
  \bibinfo{year}{2016}\natexlab{}.
\newblock \bibinfo{title}{{A Google self-driving car caused a crash for the
  first time}}.
\newblock
  \bibinfo{howpublished}{\url{https://www.theverge.com/2016/2/29/11134344/google-self-driving-car-crash-report}}.
\newblock
\newblock
\shownote{Online; accessed 18 August 2019.}


\bibitem[\protect\citeauthoryear{{The Verge}}{{The Verge}}{2019}]%
        {tesla-accident-1}
\bibfield{author}{\bibinfo{person}{{The Verge}}.}
  \bibinfo{year}{2019}\natexlab{}.
\newblock \bibinfo{title}{{Tesla hit with another lawsuit over a fatal
  Autopilot crash}}.
\newblock
  \bibinfo{howpublished}{\url{https://www.theverge.com/2019/8/1/20750715/tesla-autopilot-crash-lawsuit-wrongful-death}}.
\newblock
\newblock
\shownote{Online; accessed 18 August 2019.}


\bibitem[\protect\citeauthoryear{Tian, Pei, Jana, and Ray}{Tian
  et~al\mbox{.}}{2018}]%
        {deeptest}
\bibfield{author}{\bibinfo{person}{Yuchi Tian}, \bibinfo{person}{Kexin Pei},
  \bibinfo{person}{Suman Jana}, {and} \bibinfo{person}{Baishakhi Ray}.}
  \bibinfo{year}{2018}\natexlab{}.
\newblock \showarticletitle{DeepTest: Automated Testing of
  Deep-neural-network-driven Autonomous Cars}. In
  \bibinfo{booktitle}{\emph{Proceedings of the 40th International Conference on
  Software Engineering}} \emph{(\bibinfo{series}{ICSE '18})}.
  \bibinfo{publisher}{ACM}, \bibinfo{address}{New York, NY, USA},
  \bibinfo{pages}{303--314}.
\newblock
\showISBNx{978-1-4503-5638-1}
\urldef\tempurl%
\url{https://doi.org/10.1145/3180155.3180220}
\showDOI{\tempurl}


\bibitem[\protect\citeauthoryear{\tool}{\tool}{2019}]%
        {tool}
\tool \bibinfo{year}{2019}\natexlab{}.
\newblock \bibinfo{title}{Mis-Behaviour Prediction for Autonomous Driving
  Systems.}
\newblock
  \bibinfo{howpublished}{\url{https://github.com/icse2020submission/misbehavior-prediction/}}.
\newblock


\bibitem[\protect\citeauthoryear{track3}{track3}{2019}]%
        {mountain-track}
track3 \bibinfo{year}{2019}\natexlab{}.
\newblock \bibinfo{title}{Unity3D Snow Mountain Track.}
\newblock
  \bibinfo{howpublished}{\url{https://assetstore.unity.com/packages/3d/environments/roadways/mountain-race-track-53775}}.
\newblock


\bibitem[\protect\citeauthoryear{{Udacity}}{{Udacity}}{2017a}]%
        {udacity-simulator}
\bibfield{author}{\bibinfo{person}{{Udacity}}.}
  \bibinfo{year}{2017}\natexlab{a}.
\newblock \bibinfo{title}{{A self-driving car simulator built with Unity}}.
\newblock
  \bibinfo{howpublished}{\url{https://github.com/udacity/self-driving-car-sim}}.
\newblock
\newblock
\shownote{Online; accessed 18 August 2019.}


\bibitem[\protect\citeauthoryear{{Udacity}}{{Udacity}}{2017b}]%
        {udacity-challenge}
\bibfield{author}{\bibinfo{person}{{Udacity}}.}
  \bibinfo{year}{2017}\natexlab{b}.
\newblock \bibinfo{title}{{Udacity self-driving car's challenge}}.
\newblock
  \bibinfo{howpublished}{\url{https://github.com/udacity/self-driving-car/}}.
\newblock
\newblock
\shownote{Online; accessed 18 August 2019.}


\bibitem[\protect\citeauthoryear{{Udacity}}{{Udacity}}{2017c}]%
        {udacity-datasets}
\bibfield{author}{\bibinfo{person}{{Udacity}}.}
  \bibinfo{year}{2017}\natexlab{c}.
\newblock \bibinfo{title}{{Udacity self-driving car's datasets}}.
\newblock
  \bibinfo{howpublished}{\url{https://github.com/udacity/self-driving-car/tree/master/datasets}}.
\newblock
\newblock
\shownote{Online; accessed 18 August 2019.}


\bibitem[\protect\citeauthoryear{unity}{unity}{2019}]%
        {unity}
unity \bibinfo{year}{2019}\natexlab{}.
\newblock \bibinfo{title}{Unity3D.}
\newblock \bibinfo{howpublished}{\url{https://unity.com}}.
\newblock


\bibitem[\protect\citeauthoryear{{Vasudevan}, {Sethy}, and {Ghias}}{{Vasudevan}
  et~al\mbox{.}}{2019}]%
        {8683359}
\bibfield{author}{\bibinfo{person}{V.~T. {Vasudevan}}, \bibinfo{person}{A.
  {Sethy}}, {and} \bibinfo{person}{A.~R. {Ghias}}.}
  \bibinfo{year}{2019}\natexlab{}.
\newblock \showarticletitle{Towards Better Confidence Estimation for Neural
  Models}. In \bibinfo{booktitle}{\emph{ICASSP 2019 - 2019 IEEE International
  Conference on Acoustics, Speech and Signal Processing (ICASSP)}}.
  \bibinfo{pages}{7335--7339}.
\newblock
\showISSN{2379-190X}
\urldef\tempurl%
\url{https://doi.org/10.1109/ICASSP.2019.8683359}
\showDOI{\tempurl}


\bibitem[\protect\citeauthoryear{Zhang, Zhang, Zhang, Liu, and Khurshid}{Zhang
  et~al\mbox{.}}{2018}]%
        {deeproad}
\bibfield{author}{\bibinfo{person}{Mengshi Zhang}, \bibinfo{person}{Yuqun
  Zhang}, \bibinfo{person}{Lingming Zhang}, \bibinfo{person}{Cong Liu}, {and}
  \bibinfo{person}{Sarfraz Khurshid}.} \bibinfo{year}{2018}\natexlab{}.
\newblock \showarticletitle{DeepRoad: GAN-based Metamorphic Testing and Input
  Validation Framework for Autonomous Driving Systems}. In
  \bibinfo{booktitle}{\emph{Proceedings of the 33rd ACM/IEEE International
  Conference on Automated Software Engineering}} \emph{(\bibinfo{series}{ASE
  2018})}. \bibinfo{publisher}{ACM}, \bibinfo{address}{New York, NY, USA},
  \bibinfo{pages}{132--142}.
\newblock
\showISBNx{978-1-4503-5937-5}
\urldef\tempurl%
\url{https://doi.org/10.1145/3238147.3238187}
\showDOI{\tempurl}


\end{thebibliography}

\end{document}